\documentclass[12pt]{article}
\usepackage{amsmath,amsthm}
\usepackage{amssymb}
\usepackage{amsfonts}
\usepackage{rotating}
\usepackage{geometry}
\usepackage{booktabs}
\usepackage{tabularx}
\usepackage{multirow}
\usepackage{hyperref}
\usepackage[numbers]{natbib}

\theoremstyle{definition}

\theoremstyle{remark}

\numberwithin{equation}{section}

\begin{document}

\title{Welfare effects of information and rationality in portfolio decisions under parameter uncertainty}

\author{Michele Longo\footnote{Universit\`a Cattolica del Sacro Cuore, Largo Gemelli, 1, Milano, 20123, Italy. E-mail address: michele.longo@unicatt.it} \qquad \qquad Alessandra Mainini\footnote{Universit\`a Cattolica del Sacro Cuore, Largo Gemelli, 1, Milano, 20123, Italy. E-mail address: alessandra.mainini@unicatt.it}}
%\date{}
\maketitle

\begin{abstract}
We analyze and quantify, in a financial market with parameter uncertainty and for a Constant Relative Risk Aversion investor, the utility effects of two different boundedly rational (\textit{i.e.}, sub-optimal) investment strategies (namely, \textit{myopic} and \textit{unconditional} strategies) and compare them between each other and with the utility effect of full information. We show that effects are mainly caused by full information and predictability, being the effect of learning marginal. We also investigate the saver's decision of whether to manage her/his portfolio personally (\textit{DIY investor}) or hire, against the payment of a management fee, a professional investor and find that delegation is mainly motivated by the belief that professional advisors are, depending on investment horizon and risk aversion, either better informed (\textquotedblleft insiders\textquotedblright) or more capable of gathering and processing information rather than their ability of learning from financial data. In particular, for very short investment horizons, delegation is primarily, if not exclusively, motivated by the beliefs that professional investors are better informed.

\end{abstract}

\newpage

\section{Introduction}

Traditional portfolio management models assumed, among others, complete observability of markets' parameters and fully rational agents.
But, as soon as empirical tests revealed limits of these models to explain financial data, a vast literature has grown with the aim of
relaxing some of the classical assumptions.
We refer to Pastor and Veronesi \cite{pastor-veronesi-nber-2009} for a review of recent work on parameter uncertainty and learning in finance; whereas, reasons for incorporating bounded rationality in financial models can be found in the surveys on behavioral finance by De Bondt and Thaler \cite{debondt-thaler-95} and Barberis and Thaler \cite{barberis-thaler-2003}.

In this paper we analyze and quantify the welfare effects -- in terms of additional fraction of initial wealth that makes two investment strategies indifferent -- of two boundedly rational investment behaviors in a continuous-time financial market with parameter uncertainty,\footnote{In this paper, the expressions \textquotedblleft parameter uncertainty\textquotedblright, \textquotedblleft partial observation\textquotedblright, and \textquotedblleft partial information\textquotedblright\ are used interchangeably.} and compare them between each other and with the effect of uncertainty measured by the additional fraction of initial wealth that makes a rational agent indifferent between a full information and a partial information scenario.
In particular, we consider expected utility investors with Constant Relative Risk Aversion (CRRA) preferences over terminal wealth that trade in a continuous-time financial market with a risk-less bond and a risky asset, where the market price of risk is represented by a -- possibly unobservable -- random variable with Gaussian prior distribution.\footnote{
Since we are interested in valuing and comparing conditions such as better access to information, or better ability to process it or to behave dynamically, we abstract from transaction costs and others types of market frictions such as limited trading and assume that investors trade continuously at zero costs.}
In a dynamic context, parameter uncertainty has two effects. The first is the predictive role of observable quantities (in our setting the observable market price of risk tells about the distribution of the unobservable variable). The second effect has to do with the fact that uncertainty changes over time. Indeed, as new information comes to light through further observation, the investor updates its  posterior distribution for the unobserved parameter, thus changing the investment opportunity set which, in turn, produces an hedging demand due to the possibility of learning from these changes.
In this setting, agents may be boundedly rational in that they either do not behave dynamically (\textit{i.e.}, they do not learn from fluctuations in the conditional distribution of the unobserved parameter) but process information properly, we refer to them as \textit{myopic} investors,\footnote{Myopic strategies have also been proved to be optimal in certain optimization-based portfolio selection models such as the \textit{rational inattention} approach proposed in Huang and Liu \cite{huang-liu-2007}. Here, we are not concerned with the issue of rationalizing certain, apparently boundedly rational, behaviors but rather to evaluate and compare them.} or they also lack of competence in gathering and processing information (\textit{i.e.}, they completely disregard available information provided by prices, thus ignoring predictability of assets returns), we call them \textit{unconditional} investors.\footnote{We focus on these sub-optimal behaviors because, by citing De Bondt and Thaler \cite{debondt-thaler-95} (pp. 385-386), it is not realistic to assume that everyone in the economy is \textquotedblleft as smart as Sandy Grossman\textquotedblright, neither plausible that \textquotedblleft everyone looks toward the future in a way that would make econometricians proud\textquotedblright, which seems particularly appropriate for \textit{DIY} investors  (see Section \ref{Section-DIY-investor}).}
Our explicit results allow us to decompose the utility effect in passing from a partially informed unconditional investor (the least informed and \textquotedblleft able\textquotedblright) to a fully informed rational investor (the best informed and \textquotedblleft able\textquotedblright) into the sum of the effects of predictability, learning, and full information.
This decomposition, apparently new in this context, enables to show that welfare effects of parameter uncertainty are mainly caused,
%We find that the main effects are caused,
depending on risk aversion and investment horizon, by full information and/or predictability, being the effect of learning marginal.
In particular,  in the short run, effects are almost exclusively associated to full information and then to predictability and learning, with the value of the latter negligible compared with the previous two.
%\textbf{
%the longer the investment horizon  and/or the \textquotedblleft higher\textquotedblright\ the distance of risk aversion from the logarithmic investor, the more important is the relative contribution of predictability, whereas full information produces relevant wealth effect in the short run and for investors \textquotedblleft close\textquotedblright\ to the logarithmic one.}

Then, the analysis is adapted to investigate the saver's choice of whether to manage her/his portfolio personally (\textit{DIY} investor) or hire, at some cost, a professional investor.
Our findings support the idea that delegation is mainly motivated by the beliefs that professional investors are either better informed (\textquotedblleft insiders\textquotedblright) or more capable of gathering and processing information rather than their ability of learning from financial data.
In particular, for very short investment horizons, delegation is primarily, if not exclusively, motivated by the beliefs that professional investors are better informed.
Moreover, it is also suggested that investors with a coefficient of relative risk aversion between $0$ and $1$ are presumably the more interested in delegation.

Our work is related to the literature on continuous-time financial models with parameter uncertainty.
%which begins with Dothan and Feldman \cite{dothan-feldman-86}, Detemple \cite{detemple86}, and Gennotte \cite{gennotte-86} where, within an equilibrium framework, the Authors analyze the asset pricing implications of parameter uncertainty. Since then,
%where the issue of parameter estimation in financial models has been extensively studied in an increasing number of papers. Among others, the portfolio allocation perspective is examined in
We cite, among others, Brennan \cite{brennan98}, Lakner \cite{lakner-95}, Karatzas and Zhao \cite{karatzas-zhao-01}, Rogers \cite{rogers-01}, Xia \cite{xia01}, Honda \cite{honda-03}, Rieder and B\"auerle \cite{rieder-bauerle}, Cvitani\'{c} et al. \cite{cvitanic-lazrak-martellini-zapatero-06}, Bj\"ork et al. \cite{bjork-davis-landen-2010}, Longo and Mainini \cite{longo-mainini-2016}.
However, almost all these papers are mainly concerned with portfolio allocation effects of either bounded rationality or full information rather than with the effects on agents' utility and the few of them that consider the latter effects confine their analysis  either to special cases in terms of agent's risk aversion or to a single type of sub-optimal strategy (see, for instance, Browne and Whitt \cite{browne-whitt-1996}, Brennan and Xia \cite{brennan-xia-2010}, Cvitani\'{c} et al. \cite{cvitanic-lazrak-martellini-zapatero-06}) or just to uncertainty (see, for example, Karatzas and Zhao \cite{karatzas-zhao-01}, Brendle \cite{brendle-06}). Branger  et al. \cite{branger-larsen-munk-2013} analyze, in a more general setting, portfolio and welfare effects of learning and ambiguity in a fashion similar to the one employed in the present work.
%(see also Larsen and Munk \cite{larsen-munk-2012})

Although the way of behaving rationally is often unique within a certain paradigm, there are several ways of being boundedly rational: potentially, any sub-optimal strategy may represent a boundedly rational behavior within the model it belongs to.
In this paper we consider two types of sub-optimal policies, that is \textit{myopic} and \textit{unconditional}, but other types of sub-optimal behaviors have been explored in continuous-time portfolio management models.
For instance, sub-optimal leverage or sub-optimal diversification have been considered by Brennan and Torous \cite{brennan-torous-1999}, where, in a complete information setting, the Authors show that the cost of departing from the optimal allocation to the risky asset (\textit{sub-optimal leverage}), even of a factor as high as $.5$, is modest (and, by order of magnitude, comparable with reasonable portfolio management fees), whereas the cost of poor diversification (\textit{i.e.}, portfolios where some assets are omitted) may be much higher.
Larsen and Munk \cite{larsen-munk-2012} confirm, in a more general setting, the conclusions of Brennan and Torous \cite{brennan-torous-1999} %, in a complete observable setting,
about the welfare loss of poor diversification. They also analyze \textit{near-optimal} strategies (\textit{i.e.}, strategies obtained by small perturbations from the optimal portfolio weights) and myopic strategies, finding that losses associated to the latter investment policies are limited.
%costs suffered by long-term power investors by deviating from the optimal investment strategy; in particular, the work focuses on
%sub-optimal strategies such as those obtained by omitting some assets from the portfolio (poor diversification as in Brennan and Torous \cite{brennan-torous-1999}), the myopic strategy and, in general, \textit{near-optimal} strategies (\textit{i.e.}, strategies obtained by small perturbations from the optimal portfolio weights).
%
%Another important sub-optimal behavior is, according to
Rogers \cite{rogers-01} analyzes the sub-optimal strategy named \textit{relaxed}: that is, the investor changes portfolio allocations at fixed times (the extreme case being the \textit{buy-and-hold} strategy).
%, where the investor chooses a number of shares and bonds at the beginning of the investment period and keeps them for the whole period without further trading).
The Author provides, in a completely observable model, a power expansion of the corresponding cost
%for the cost of being \textit{relaxed}, as a function of the time lag,
and shows that it is not so high, at least if compared with the cost of uncertainty.
%
%From a mathematical point of view,
In Haugh et al. \cite{haugh-kogan-wang-2006}, the Authors develop an algorithm to evaluate sub-optimal policies in portfolio problems with trading constraints by means of bounds on the utility loss associated to a given sub-optimal strategy and apply the algorithm to evaluate the quality of unconditional (that they call \textit{static}) and myopic strategies in the context of return predictability.

The paper is organized as follows.
In Section \ref{section-investment-problem}, we describe the investment problem and solve it under several specifications about the investor rationality and/or knowledge of market parameters.
In particular, in Section  \ref{Section-informed}, we assume that the investor knows the distribution of the market price of risk and invests accordingly.
Then, we assume that the investor does not observe the market price of risk % but knows only its prior
and determine explicitly the expected utility for a rational (Section  \ref{Section-learner}), a myopic (Section \ref{Subsection-myopic}) and an unconditional (Section \ref{Subsection-static}) agent.
In Section \ref{Section-comparative-analysis}, we compare the scenarios studied in Section \ref{section-investment-problem} and decompose additively the whole wealth effect into the sum of the single effects. We also provide simulations that give insights about the importance of each effect.
Section \ref{Section-DIY-investor} investigates the saver's decision of whether to manage her/his portfolio personally (\textit{DIY} investor) or to delegate its management to a professional investor.
Section \ref{Section-Conclusion} concludes.

\section{The investment problem} \label{section-investment-problem}
In a continuous-time frictionless financial market, an investor with investment horizon $T>0$ and initial wealth $x > 0$ trades two assets: a risk-free asset (bond) with constant interest rate $r\geq0$ and a risky asset (stock) whose price process $\left(S_t\right)_{t \geq 0}$ satisfies the equation
\begin{equation}
dS_t=S_t\left(\left(r+\sigma\Theta\right)dt+\sigma dW_t\right),  \label{stock-price-S}
\end{equation}
where $\left(W_t\right)_{t \geq 0}$ is a standard Brownian motion, $\sigma $  is a positive constant, and $\Theta$, the \textit{market price of risk},  is a random variable independent of $(W_t)_{t \geq 0}$ and with known prior distribution normally distributed with mean $\theta_0$ and variance $v_0$. If the investor chooses the trading strategy $\pi=\left(\pi(t)\right)_{t \in [0,T]}$, where $\pi(t) \in \mathbb R$ represents the fraction of wealth invested in the risky asset at time $t$, then his/her wealth $\left(X_t\right)_{t \geq 0}$ %$\left(X_t\right)_{t \in [0,T]}$
evolves according to the budget equation
\begin{equation}
dX_t=rX_tdt+\sigma\pi(t)X_t(\Theta dt +dW_t), \;\;\; X_0=x,\label{wealth}
\end{equation}
and we assume that he/she enjoys expected utility (from terminal wealth)
\begin{equation}
\mathbb{E}\left[u_{\gamma}\left(X_T\right)\right], \label{expected-value-original}
\end{equation}
where
\begin{equation}
u_{\gamma }\left( x\right) :=\left\{
\begin{array}{l}
\displaystyle{\frac{x^{1-\gamma }}{1-\gamma }},\;\;\;\;\gamma >0,\;\;\gamma \neq 1, \medskip \\
\ln x,\;\;\;\;\;\;\;\,\gamma =1,%
\end{array}%
\right.
\end{equation}
being $\gamma$ the investor's  coefficient  of  relative  risk  aversion. Investors with $\gamma > 1$ are known in literature as \textit{conservative}, whereas those with risk aversion $0 < \gamma < 1$ are called \textit{aggressive}.\footnote{Empirical evidence suggests values for $\gamma$ between $1$ and $10$.}
In order to avoid infinite expected utility,  we assume throughout the paper that parameters satisfy the condition $\gamma(1 + v_0 T)- v_0T>0$.
Notice that: (i) if we fix $v_0>0$ and $\gamma >0 $, then, if $\gamma \geq 1$, $T$ can range in the entire positive half-line, whereas, if $0 < \gamma < 1$, the condition holds for $T<\bar{T}:=\gamma /\left( v_{0}\left( 1-\gamma \right) \right) $; if we fix $v_0>0$ and $T\geq 0$, then we consider all $\gamma > \bar{\gamma}:=v_0 T/(1+v_0 T)$ (notice that $0<\bar{\gamma}<1$).

We now determine explicitly the expected utility in (\ref{expected-value-original}) under several specifications about agents' knowledge of $\Theta$ and/or agents' rationality. In particular, we first assume that the investor knows the true distribution of $\Theta$ and invests accordingly (Section \ref{Section-informed}).
Then, we consider a setting where the investor does not observe $\Theta$ but knows only its prior, and calculate the expected utilities of a rational (\textit{i.e.}, an investor that, by observing stock returns, continuously updates her/his beliefs about the unobserved market price of risk, Section  \ref{Section-learner}), a myopic (\textit{i.e.}, an investor that properly learns about the unobservable market price of risk by observing stock returns, but ignores the dynamic aspect of the investment decision, Section \ref{Subsection-myopic}) and an unconditional (\textit{i.e.}, an investor that completely disregards available information provided by the observed market price of risk and sticks to its prior, Section \ref{Subsection-static}) agent.

\subsection{Fully informed investors} \label{Section-informed}
A \textit{fully informed} and rational investor ($I$) observes $\Theta$ as well as $\left(W_t\right)_{t \geq 0}$ and maximizes the expected value (\ref{expected-value-original}), subject to the wealth dynamics (\ref{wealth}), over all $(\mathcal{F}^{\Theta, W}_t)$-progressively measurable (and integrable) investment strategies $\pi$, where $(\mathcal{F}^{\Theta, W}_t)_{t\geq 0}$ is the filtration generated by $\Theta$ and $\left(W_t\right)_{t \geq 0}$.
The agent invests according to the standard Merton's strategy
\begin{equation}
\pi _{\gamma }^{I}(t)=\frac{\Theta}{\sigma \gamma},\;\;\; 0 \leq t \leq T, \;\;\; \gamma >0,\label{optimal-portfolio-informed-power}
\end{equation}
and enjoys the expected utility (see Appendix \ref{appendix-informed})
\begin{equation}
 V^{I}(x;\gamma,T) =\left\{
\begin{array}{l}
\displaystyle{\frac{x^{1-\gamma}}{1-\gamma}}\exp\left(\varphi^I(\gamma,T)\theta_0^2 +\psi^I(\gamma,T)\right), \;\;\ \gamma >0, \; \gamma \neq 1,  \medskip \\
\displaystyle{\ln x+rT+\frac{\theta_0^2T}{2}+\frac{v_0T}{2} ,\;\;\;\ \gamma =1},%
\end{array}%
\right. \label{utilita-indiretta-informed-power}
\end{equation}
where, for $T\geq 0$, $\gamma >0$ and $ \gamma \neq 1$,
\begin{equation}
\varphi^I(\gamma,T) : =  \frac{(1 - \gamma)T}{2(\gamma(1 + v_0 T)- v_0T)}
\end{equation}
and
\begin{equation}
\psi^I(\gamma,T) : = \frac{1}{2}\ln \left(\frac{\gamma}{\gamma(1 + v_0 T)- v_0T}\right)+ r(1 - \gamma)T .
\end{equation}

\subsection{Partially informed investors} \label{Section-partial-information}
In this section we assume that investors do not observe $\Theta$ nor the Brownian motion $\left(W_t\right)_{t \geq 0}$, whereas they know the prior of $\Theta$ and can continuously observe $(S_t)_{t \geq 0}$.
The resulting portfolio problem, which is of \textit{partial information} type, has been extensively studied in literature (see, among others, Karatzas and Zhao \cite{karatzas-zhao-01}, Rogers \cite{rogers-01}, Rieder and B\"auerle \cite{rieder-bauerle}, Bj\"ork et al. \cite{bjork-davis-landen-2010}, and Pham \cite{pham-2011}).
Typically, the problem is first reduced, by means of filtering techniques, within a complete observation setting and then solved either by martingale methods or by dynamic programming techniques.
Since the analysis of such problems is by now standard, we just recall the definitions we will use later to represent different investment strategies and refer to the literature for details (see, for instance, Rishel \cite{rishel99}, Karatzas and Zhao \cite{karatzas-zhao-01},  or Rogers \cite{rogers-01}, Section 6).
Let $\left(\mathcal{F}^S_t\right)_{t \geq 0}$ be the filtration generated by $\left(S_t\right)_{t \geq 0}$ and define the process
\begin{equation}
dY_t=\hat{\Theta}(t,Y_t)dt+d\hat W_t, \;\;\; Y_0=0,  \label{process_Y-2}
\end{equation}
where
\begin{equation}
\hat{\Theta}(t,y):=\frac{\theta_0+v_0y}{1+v_0t} \label{conditional-mean}
\end{equation}
and $(\hat W_t)_{t \geq 0}$ is a Brownian motion w.r.t. the observable filtration $\left(\mathcal{F}^S_t\right)_{t \geq 0}$.
Then
\begin{equation} \label{conditional-mean-2}
 \hat{\Theta}(t,Y_t)= \mathbb E[\Theta\mid\mathcal F_t^S],  \smallskip \;\;\;\;\;\;\;\; 0 \leq t \leq T,
\end{equation}
(\textit{i.e.}, $\hat{\Theta}(t,Y_t)$ is the time $t$ Bayesian estimation of $\Theta$, conditional on the available information $\mathcal{F}^S_t$) and
\begin{equation}  \label{wealth2}
dX_t=rX_tdt+\sigma\pi(t) X_t(\hat{\Theta}(t,Y_t)dt +d\hat W_t), \;\;\; X_0=x,
\end{equation}
represents the investor's budget constraint in terms of observable quantities, where $\pi$ is a $\left(\mathcal{F}^S_t\right)$-progressively measurable (and integrable) investment strategy.

\subsubsection{Rational investors} \label{Section-learner}

In a dynamic context under partial information, a \textit{rational investor} ($R$) properly updates his/her beliefs about the unobservable parameter $\Theta$, conditional on the observed market price of risk,
and hedges against future changes in the investment opportunity set, when building his/her portfolio, by learning from fluctuations in the filtered marked price of risk.
Formally, such an investor maximizes (\ref{expected-value-original}) over all $\left(\mathcal{F}^S_t\right)$-progressively measurable (and integrable) investment strategies $\pi $, under the two-dimensional state dynamics (\ref{process_Y-2}), (\ref{wealth2}).
This is a \textit{complete observation} and Markovian control problem that is solved by using dynamic programming techniques  (see Appendix \ref{appendix-rational}).
The optimal investment strategy and the expected utility are, respectively,
\begin{equation}
\pi _{\gamma}^{R}(t)=\frac{\hat{\Theta}(t,Y_{t})}{\sigma \gamma }+\frac{\hat{\Theta}(t,Y_{t})}{\sigma \gamma }\left( \frac{(1-\gamma )(T-t)v_{0}}{\gamma (1+v_{0}T)-v_{0}T+v_{0}t}\right),\;\;\; 0 \leq t \leq T, \;\;\; \gamma >0,          \label{optimal-portfolio-learner-power}         \end{equation}
and
\begin{equation}
 V^{R}(x;\gamma,T) =\left\{
\begin{array}{l}
\displaystyle{\frac{x^{1-\gamma }}{1-\gamma }\exp \left( \varphi ^{R}\left( \gamma ,T\right) \theta _{0}^{2}+\psi ^{R}\left( \gamma ,T\right) \right), \;\;\;\ \gamma >0, \; \gamma \neq 1},  \medskip \\
\displaystyle{\ln x+rT+\frac{\theta_0^2T}{2}+\frac{v_0T}{2}-\frac{1}{2}\ln(1+v_0 T),\;\;\;\ \gamma =1},%
\end{array}%
\right. \label{utilita-indiretta-learner-power}
\end{equation}
where, for $T\geq 0$, $\gamma >0$ and $ \gamma \neq 1$,
\begin{equation}
\varphi^R(\gamma,T) : =  \frac{(1 - \gamma)T}{2(\gamma(1 + v_0 T)- v_0T)}
\end{equation}
and
\begin{equation}
\psi^R(\gamma,T) : = \frac{1}{2}\left( \gamma \ln \frac{\gamma }{\gamma (1+v_{0}T)-v_{0}T}-(1-\gamma)\ln \left( 1+v_{0}T\right) \right)+ r\left( 1-\gamma \right) T.
\end{equation}

\subsubsection{Myopic investors} \label{Subsection-myopic}

A \textit{myopic investor} ($M$) properly updates, by means of the Bayesian estimator (\ref{conditional-mean-2}), the conditional distribution of $\Theta$ but ignores the hedging demand associated to fluctuations on the filtered market price of risk (that is, he/she considers the estimation as it were constant from $t$ onwards).
Hence, he/she chooses the investment policy\footnote{Notice that, according to Kuwana \cite{kuwana-95}, the logarithmic investor is myopic, that is $\pi _{1}^{M}(t) = \pi _{1 }^{R}(t)$, for all $t\geq 0$.}  %\ul{FIN QUI}
\begin{equation}
\pi _{\gamma}^{M}(t) =\frac{\hat{\Theta}(t,Y_{t})}{\sigma \gamma},\;\;\; 0 \leq t \leq T, \;\;\; \gamma >0,\label{optimal-portfolio-myopic-power}
\end{equation}
and enjoys an expected utility $V^{M}$, given by (\ref{expected-value-original}), where the wealth process evolves according to (\ref{wealth2}) with $\pi =\pi_{\gamma}^{M}$.
After some algebra  (see Appendix \ref{appendix-myopic}) one obtains:
\begin{equation}
 V^{M}(x;\gamma,T) =\left\{
\begin{array}{l}
\displaystyle{\frac{x^{1-\gamma}}{1-\gamma}\exp\left(\varphi^M(\gamma,T)\theta_0^2 + \psi^M(\gamma,T) \right), \;\;\;\ \gamma >0, \; \gamma \neq 1},  \medskip \\
\displaystyle{\ln x+rT+\frac{\theta_0^2T}{2}+\frac{v_0T}{2}-\frac{1}{2}\ln(1+v_0 T),\;\;\;\ \gamma =1},%
\end{array}%
\right. \label{utilita-indiretta-myopic-power}
\end{equation}
where, for $T\geq 0$, $\gamma >0$ and $ \gamma \neq 1$,
\begin{equation}
\varphi^M(\gamma,T) : = \frac{(1-\gamma )\left( (1+v_{0}T)^{r_{2}-r_{1}}-1\right) }{2\gamma v_{0}\left( r_{2}(1+v_{0}T)^{r_{2}-r_{1}}-r_{1}\right) },
\end{equation}
\begin{equation}
\psi^M(\gamma,T) : = \frac{1}{2}\left( \ln \frac{r_{2}-r_{1}}{r_{2}(1+v_{0}T)^{r_{2}-r_{1}}-r_{1}}+r_{2}\ln (1+v_{0}T)\right) + r(1 - \gamma)T,
\end{equation}
and $r_1 , r_2$, with $r_1 < r_2$, are the solutions of the quadratic equation:
\begin{equation}
r^2 + \left(\frac{2}{\gamma} -1 \right)r + \frac{1 - \gamma}{\gamma} =0.
\end{equation}

\subsubsection{Unconditional investors} \label{Subsection-static}

Here the degree of bounded rationality is even more severe in that an \textit{unconditional investor} ($U$) do not even update the conditional distribution of $\Theta$ and considers the market price of risk constant and equal to its unconditional mean $\theta_0=\mathbb{E}[\Theta]$ for the entire investment period, thus ignoring both predictability of and learning from the available information provided by stock returns. Notice that unconditional investors are also myopic because there is no possibility of learning since the investment opportunity set is constant.
The unconditional strategy takes the form (cf. Haugh et al. \cite{haugh-kogan-wang-2006} or Brennan and Xia \cite{brennan-xia-2010})
\begin{equation}
\pi _{\gamma}^{U}(t)=\frac{\theta_0}{\sigma \gamma},\;\;\; 0 \leq t \leq T, \;\;\; \gamma >0,  \label{optimal-portfolio-static-power}
\end{equation}
and the corresponding expected utility, $V^{U}$, is given by (\ref{expected-value-original}), where $X_T$ is determined via the wealth dynamics (\ref{wealth2}) with $\pi =\pi_{\gamma}^{U}$.
After some algebra  (see Appendix \ref{appendix-unconditional}), we get:
\begin{equation}
 V^{U}(x;\gamma,T) =\left\{
\begin{array}{l}
\displaystyle{\frac{x^{1-\gamma}}{1-\gamma}\exp\left(\varphi^U(\gamma,T)\theta_0^2 + \psi^U(\gamma,T)  \right), \;\;\;\ \gamma >0, \; \gamma \neq 1},  \medskip \\
\displaystyle{\ln x+rT+\frac{\theta_0^2T}{2},\;\;\;\ \gamma =1},%
\end{array}%
\right. \label{utilita-indiretta-static-power}
\end{equation}
where, for $T\geq 0$, $\gamma >0$ and $ \gamma \neq 1$,
\begin{equation}
\varphi^U(\gamma,T) : =  \frac{(1 - \gamma)(\gamma T + (1 - \gamma)v_0T^{2})}{2\gamma^2}
\end{equation}
and
\begin{equation}
\psi^U(\gamma,T) : =  r(1 - \gamma)T.
\end{equation}

\section{Value of full information, predictability and learning} \label{Section-comparative-analysis}

We compare the scenarios studied in the previous sections and quantify the difference between any two of them in terms of additional fraction of initial wealth needed to obtain the same expected utility.\footnote{A similar analysis, though with a different cost definition, and just for the logarithmic utility and without considering the myopic strategy, is performed in Browne and Whitt \cite{browne-whitt-1996}.}
By observing that, for all $x>0$, $\gamma >0$ and $T>0$,\footnote{Moreover, for each $i$, the expected utility of the logarithmic investor $V^{i}(x;1,T)$ is the limit of $V^{i}(x;\gamma,T)-1/(1-\gamma) $, as $\gamma \to 1$, for any $x,T>0$.}
\begin{equation}
V^{U}(x;\gamma, T) \leq V^{M}(x;\gamma, T) \leq  V^{R}(x;\gamma, T) \leq V^{I}(x;\gamma, T),
\end{equation}
we introduce on the set $\left\{ I,R,M,U\right\}$, where $I,R,M$ and $U$ refer, respectively, to the scenarios analyzed in Sections  \ref{Section-informed}, \ref{Section-learner}, \ref{Subsection-myopic} and \ref{Subsection-static},  the order relation $U\precsim M\precsim R\precsim I$, and, for $i,j \in \left\{ I,R,M,U\right\} $, with $i\precsim j$, we define $C^{ij}$ as the quantity such that
\begin{equation}
V^{i}(x;\gamma, T) = V^{j}(x(1-C^{ij});\gamma, T). \label{cost-definition-implicit}
\end{equation}
Since the passage from scenario $i$ to scenario $j$, with $i\precsim j$, is valuable and, possibly, onerous, $C^{ij}$ may be seen as the maximum \textit{cost},\footnote{Our definition of cost is identical to the notion of \textit{utility loss} defined in Branger et al \cite{branger-larsen-munk-2013}. It is also similar to the concept of \textit{wealth-equivalent utility loss} employed in Larsen and Munk \cite{larsen-munk-2012}. Moreover, $1- C^{ij}$ is called \textit{efficiency} in Rogers \cite{rogers-01} and its inverse is the ratio of the \textit{certainty equivalent} in Brennan and Xia \cite{brennan-xia-2010}, Section 19.4.
%See also Brendle \cite{brendle-06}, Section 5.
} in terms of fraction of initial wealth,
that an investor acting in scenario $i$ is willing to pay to move to scenario $j$ (a \textit{reservation price}).
In particular, in a dynamic context with parameter uncertainty, $C^{UM}$ may be considered a measure (in terms of additional fraction of initial wealth) of the value of the sole predictability (\textit{i.e.}, the value of using the conditional distribution for the unobserved market price of risk), $C^{MR}$ a measure of the value of learning (\textit{i.e.}, the value of behaving dynamically), once predictability has been considered, $C^{RI}$ a measure of the value of full information once rationality (\textit{i.e.}, predictability and learning) has been considered,
and $C^{UI}$ a measure of the benefit of considering the three aspects of portfolio decision all together.
Notice that $C^{ij}$ represents the cost over the entire investment horizon -- a sort of cumulated or compound cost --, in Section \ref{Section-DIY-investor} we introduce an annual version of the cost.

Thanks to the explicit form of expected utilities $V^{I}$, $V^{R}$, $V^{M}$ and $V^{U}$, we have:
\begin{equation}
C^{ij}=C^{ij}\left( \gamma ,T\right) =1-\exp \left( \frac{ \varphi ^{i}\left( \gamma ,T\right) -\varphi ^{j}\left( \gamma ,T\right) }{1-\gamma }\theta _{0}^{2} + \frac{\psi ^{i}\left( \gamma ,T\right) -\psi ^{j}\left( \gamma,T\right) }{1-\gamma }
\right),  \label{cumulated-costs-definition}
\end{equation}
$i,j\in \left\{ I,R,M,U\right\} $, $\gamma >0, \gamma \neq 1$.
Costs are independent of the initial wealth $x$ and logarithmic costs (\textit{i.e.}, $\gamma=1$), which have the following simple form:
\begin{eqnarray}
& & C^{UM}\left( 1,T\right)  = 1-e^{-v_{0}T/2}\left( 1+v_{0}T\right) ^{1/2}, \label{C-log-UM} \\
& & C^{MR}\left( 1,T\right)  = 0, \label{C-log-MR} \\
& & C^{RI}\left( 1,T\right)  = 1-\left( 1+v_{0}T\right) ^{-1/2}, \label{C-log-RI}
\end{eqnarray}
are equal to the limits, as $\gamma \rightarrow 1$, of the corresponding power costs.

Representation (\ref{cumulated-costs-definition}) allows to write the identity
\begin{equation}
C^{UI} = 1-\left( 1-C^{UM}\right) \left( 1-C^{MR}\right) \left( 1-C^{RI}\right), \label{cost-decomposition}
\end{equation}
which implies, at least when costs are sufficiently low -- and this is true for significant ranges of parameters $\gamma$ and $T$, and for reasonable values of the other parameters --, the following approximation:
\begin{equation}
C^{UI}\approx C^{UM}+C^{MR}+ C^{RI}. \label{cost-approximation}
\end{equation}
That is, the cost in passing from a partially informed unconditional investor (the least informed and \textquotedblleft able\textquotedblright) to a fully informed rational investor (the best informed and \textquotedblleft able\textquotedblright) is split into the sum of the costs of predictability, learning, and full information.
Notice that (\ref{cost-decomposition}) and (\ref{cost-approximation}) hold true also for the logarithmic case (\textit{i.e.}, $\gamma =1$).

\begin{figure}
\caption{Cumulated costs against investment horizon $T \,(\leq 250)$ for $\sigma=0.202$, $\theta_0 = 0.08/\sigma$ and $v_0=(0.0243/\sigma)^2$.}
\includegraphics[width=1\columnwidth]{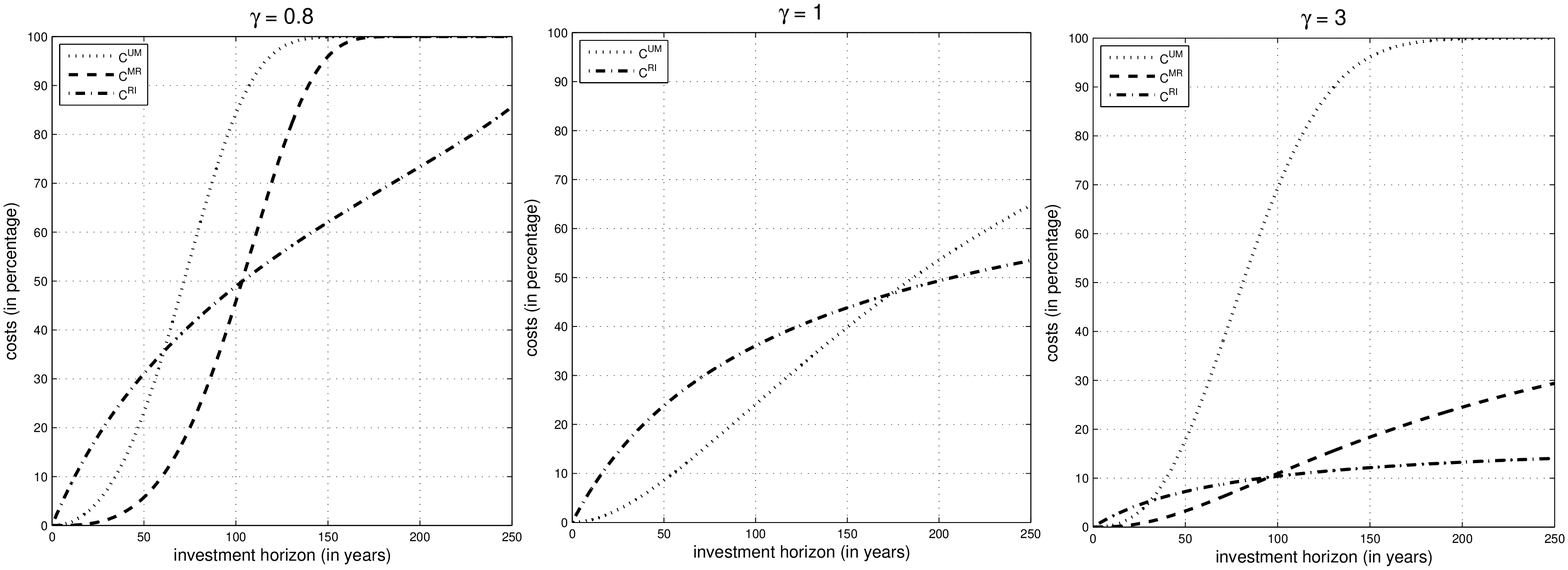}\\
\includegraphics[width=1\columnwidth]{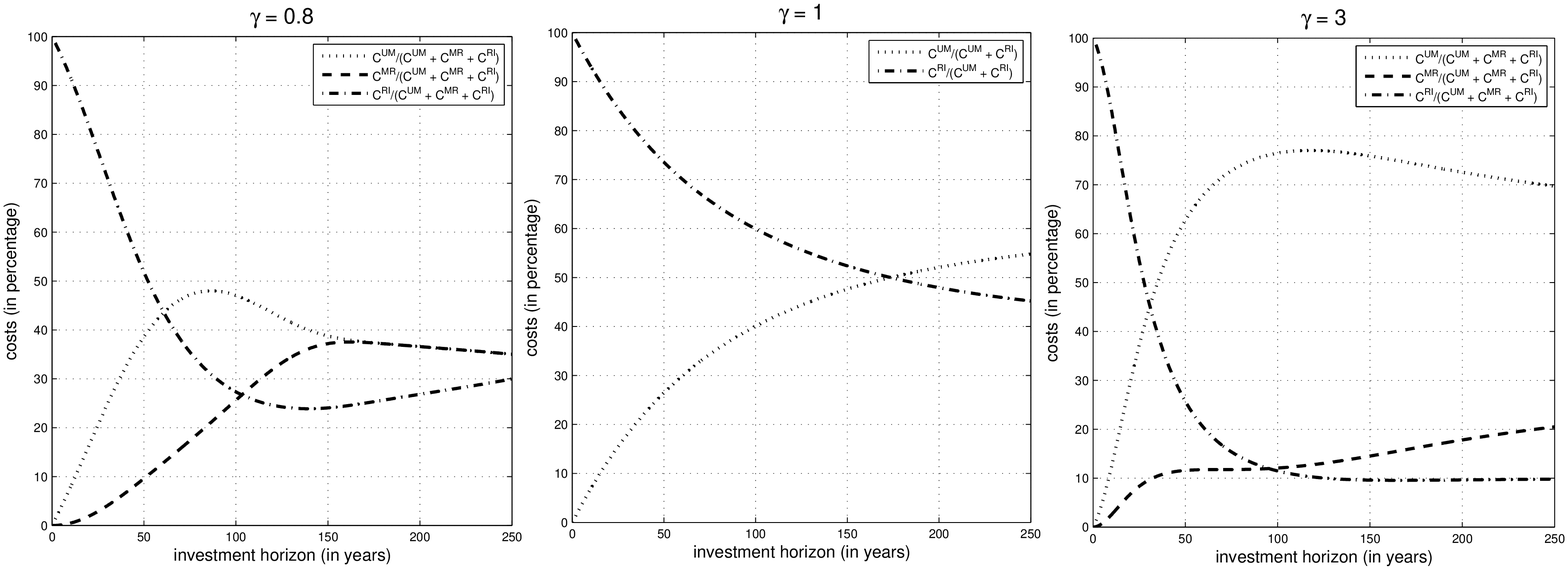}
 \label{figura_costi_rispetto_T_lungo}
\end{figure}

The rest of the section is devoted to the analysis of
$ C^{UM}$, $C^{MR}$, $ C^{RI}$ %and $C^{UI} $ (as well as
and the goodness of approximation (\ref{cost-approximation}) with respect to the investment horizon $T$ and the risk aversion $\gamma$. We start from the investment horizon. Standard calculus yields:\footnote{The notation $f\left( x\right) =o\left( g\left( x\right) \right) $, as $x\rightarrow x_{0}$, means, as usual, $\lim_{x\rightarrow x_{0}}f\left( x\right) /g\left( x\right) =0$.}
\begin{eqnarray}
& &C^{UM}\left( \gamma ,T\right)  =\frac{v_{0}^{2}}{4\gamma }T^{2}+o\left(T^{2}\right) ,\;\;\;T\rightarrow 0, \\
& &C^{MR}\left( \gamma ,T\right)  =o\left( T^{2}\right) ,\;\;\;T\rightarrow 0,\\
& &C^{RI}\left( \gamma ,T\right)  =\frac{v_{0}}{2\gamma }T+\frac{v_{0}^{2}}{4\gamma }\left( \frac{1}{2\gamma }-2\right) T^{2}+o\left( T^{2}\right)
,\;\;\;T\rightarrow 0, \\
& &C^{UI}\left( \gamma ,T\right)  =\frac{v_{0}}{2\gamma }T+\frac{v_{0}^{2}}{4\gamma }\left( \frac{1}{2\gamma }-1\right) T^{2}+o\left( T^{2}\right)
,\;\;\;T\rightarrow 0,
\end{eqnarray}
for each $\gamma > 0$.
Hence, all costs vanish as $T$ shrinks to $0$ and, by observing that
\begin{equation} \label{cost-approximation-analitically}
C^{UI}\left( \gamma ,T\right) =C^{UM}\left( \gamma ,T\right) +C^{MR}\left(\gamma ,T\right) +C^{RI}\left( \gamma ,T\right) +E\left( \gamma ,T\right),%
\end{equation}
where the \textit{error} $E\left( \gamma ,T\right)$ is such that
\begin{equation} \label{cost-approximation-analitically-error}
E\left( \gamma ,T\right)= o\left( T^{2}\right),\;\;\;T\rightarrow 0,
\end{equation}
we expect approximation (\ref{cost-approximation}) to be extremely accurate for sufficiently short investment horizons, as simulations ahead in the section confirm.
We also notice that, in the short run, costs are mainly motivated by full information and then by predictability and learning, with the value of the latter negligible compared with the previous two.
Reasonably, predictability and learning display their effects in the long run.
Costs' behavior as the investment horizon gets longer depends on $\gamma$. We distinguish three cases: \textit{(i)} $\gamma > 1$, costs exist for all $T\geq 0$ and \begin{equation}
\lim_{T\rightarrow \infty }~C^{UM}\left( \gamma ,T\right) =\lim_{T\rightarrow \infty }~C^{MR}\left( \gamma ,T\right) =1, \;\;\;
\lim_{T\rightarrow \infty }~C^{RI}\left( \gamma ,T\right) =1-\sqrt{1-1/\gamma};%=1-\sqrt{\frac{\gamma -1}{\gamma }};
\end{equation}
\textit{(ii)} $\gamma = 1$ (the logarithmic case, see (\ref{C-log-UM})-(\ref{C-log-RI})), costs exist for all $T\geq 0$, $C^{MR}\left( 1 ,T\right)=0$, for all $T$, and
\begin{equation}
\lim_{T\rightarrow \infty }~C^{UM}\left( 1 ,T\right) =\lim_{T\rightarrow \infty }~C^{RI}\left( 1 ,T\right) =1;
\end{equation}
\textit{(iii)} $0 < \gamma <1$, costs exist for all $T$ not exceeding $\bar{T}$, and %$\gamma / (v_0(1-\gamma))=:\bar{T}$
\begin{equation}
\lim_{T\rightarrow \bar{T}}~C^{UM}\left( \gamma ,T\right) = C^{UM}\left( \gamma ,\bar{T}\right)<1, \;\;
\lim_{T\rightarrow \bar{T}}~C^{MR}\left( \gamma ,T\right) =
\lim_{T\rightarrow \bar{T}}~C^{RI}\left( \gamma ,T\right) = 1.
\end{equation}
Notice that for very long investment horizons -- though financially implausible -- costs, except $C^{MR}\left( 1 ,T\right)$, which is identically $0$, become relevant (see Figure \ref{figura_costi_rispetto_T_lungo}) and, as a consequence, approximation (\ref{cost-approximation}) does not hold. Nevertheless, for financially reasonable values of the parameters, our simulations show that the approximation applies well for realistic investment horizons such as those not exceeding $30$ years (see Table \ref{table-0243}).
Lastly, since costs refer to the entire investment horizon $T$, we expect them to increase with respect to $T$ and, indeed, it can be proved (though, with some algebra) that, for all $\gamma >1$, $C^{UM}$, $C^{MR}$ and $C^{RI}$ are increasing with respect to $T$. $C^{UM}$ and $C^{RI}$ are increasing also for $\gamma=1$ (see (\ref{C-log-UM})-(\ref{C-log-RI})).

We now turn to costs' behavior with respect to the coefficient of relative risk aversion $\gamma$. We recall that, for any fixed $T>0$, costs exist for all $\gamma > \bar{\gamma}=v_0 T/(1+v_0 T)$. We have
\begin{equation}
\lim_{\gamma \rightarrow \infty }~C^{UM}\left( \gamma ,T\right) = \lim_{\gamma \rightarrow \infty }~C^{MR}\left( \gamma ,T\right) =
\lim_{\gamma \rightarrow \infty }~C^{RI}\left( \gamma ,T\right) = 0,
\end{equation}
that is, for $\gamma$ sufficiently high, costs become very small, approximation (\ref{cost-approximation}) is accurate and, indeed, we also have $C^{UI}\rightarrow 0$ as $\gamma\rightarrow \infty$.
All costs get smaller as risk aversion increases because investment in the risky asset (see, expressions (\ref{optimal-portfolio-informed-power}), (\ref{optimal-portfolio-learner-power}), (\ref{optimal-portfolio-myopic-power}) and (\ref{optimal-portfolio-static-power})), the only one that carries uncertainty, approach to $0$ as $\gamma$ increases to infinity, hence, information affects a very little part of the investor's wealth and, consequently, its value becomes negligible. As already noticed, costs tend to the logarithmic case as $\gamma$ approaches to $1$, and, as $\gamma \rightarrow \bar{\gamma}$, we have
\begin{equation}
\lim_{\gamma \rightarrow \bar{\gamma}}~C^{UM}\left( \gamma ,T\right) = C^{UM}\left( \bar{\gamma} ,T\right)<1, \;
\lim_{\gamma\rightarrow \bar{\gamma}}~C^{MR}\left( \gamma ,T\right) =
\lim_{\gamma\rightarrow \bar{\gamma}}~C^{RI}\left( \gamma ,T\right) = 1.
\end{equation}
Finally, $C^{RI}(\gamma, T)$ is decreasing with respect to $\gamma$ for all $T>0$, whereas simulations ahead show that other costs are not monotonic in general.

\begin{figure}
\caption{Cumulated costs against investment horizon $T \,(\leq 30)$ for $\sigma=0.202$, $\theta_0 = 0.08/\sigma$ and $v_0=(0.0243/\sigma)^2$.}
\includegraphics[width=1\columnwidth]{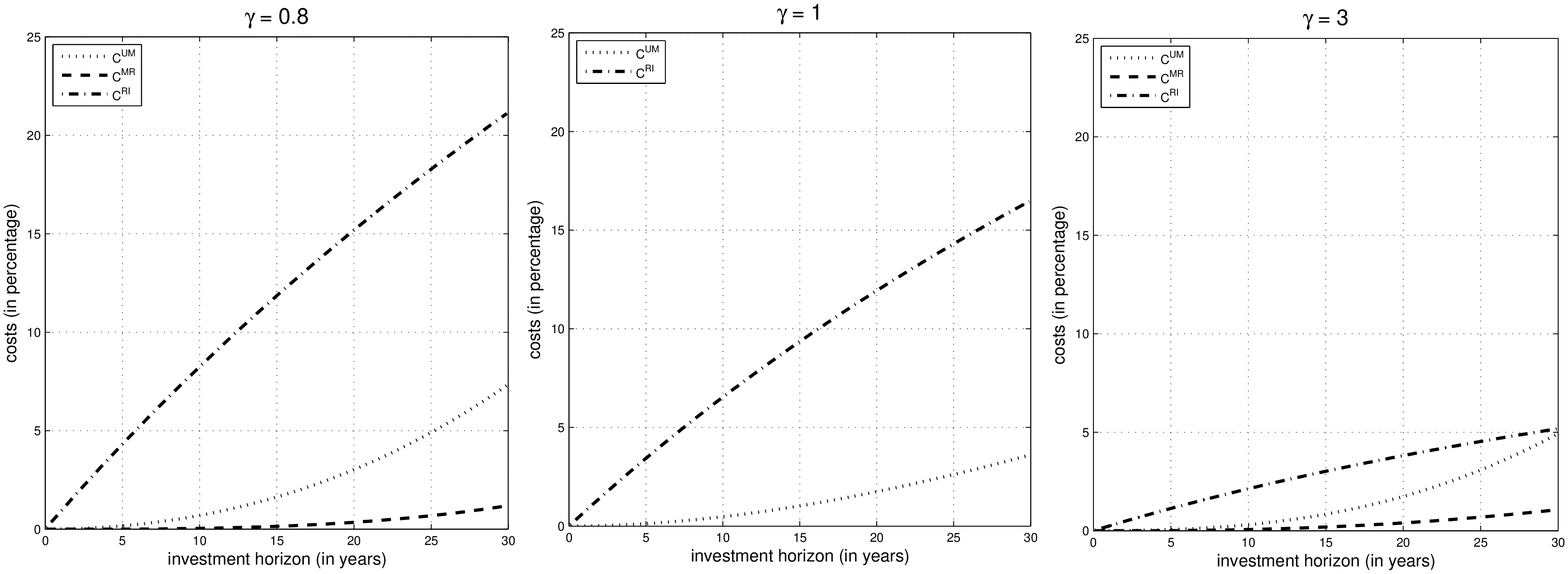}\\
\includegraphics[width=1\columnwidth]{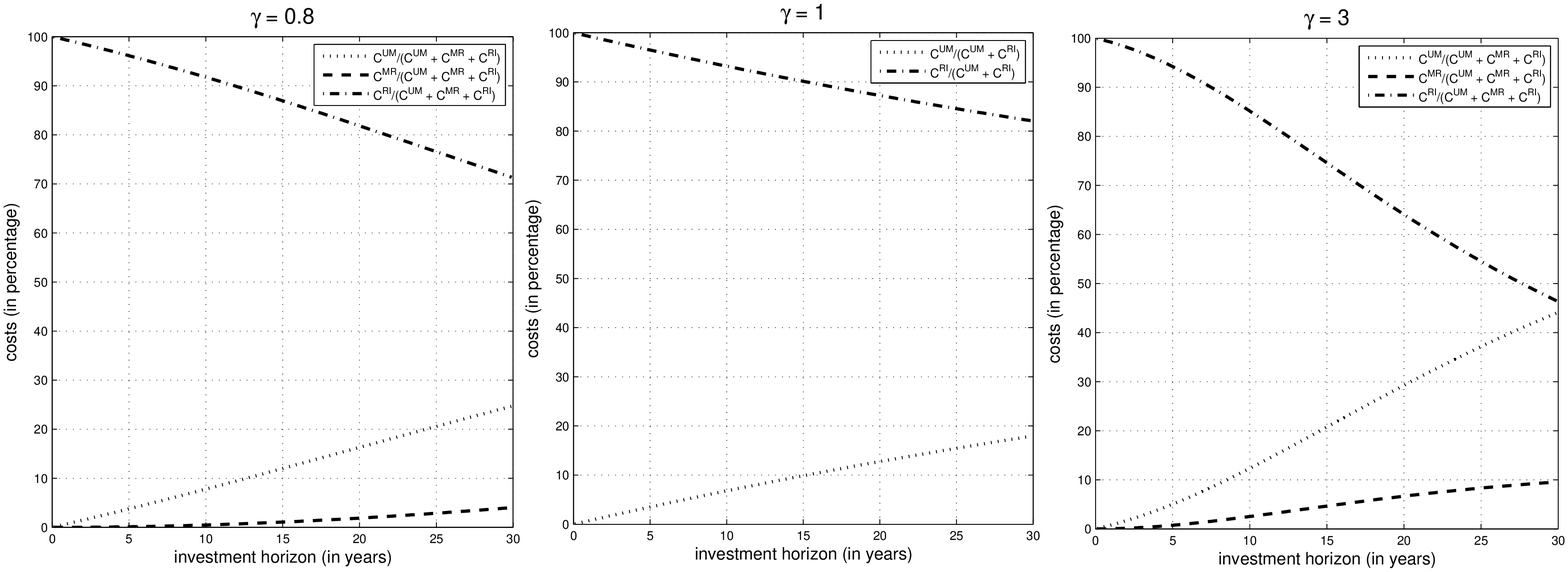}
 \label{figura_costi_rispetto_T}
\end{figure}

\begin{figure}
\caption{Cumulated costs against risk aversion $\gamma \,(\leq 12)$ for $\sigma=0.202$, $\theta_0 = 0.08/\sigma$ and $v_0=(0.0243/\sigma)^2$.}
\includegraphics[width=1\columnwidth]{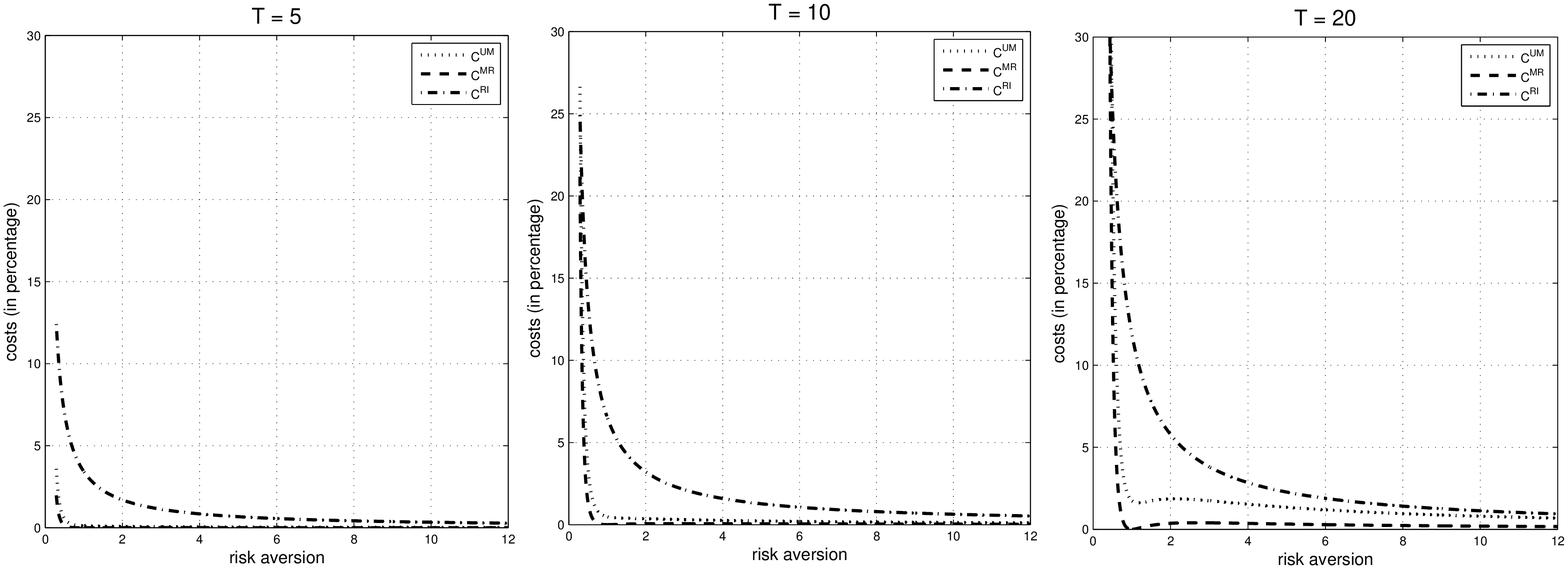} \\
\includegraphics[width=1\columnwidth]{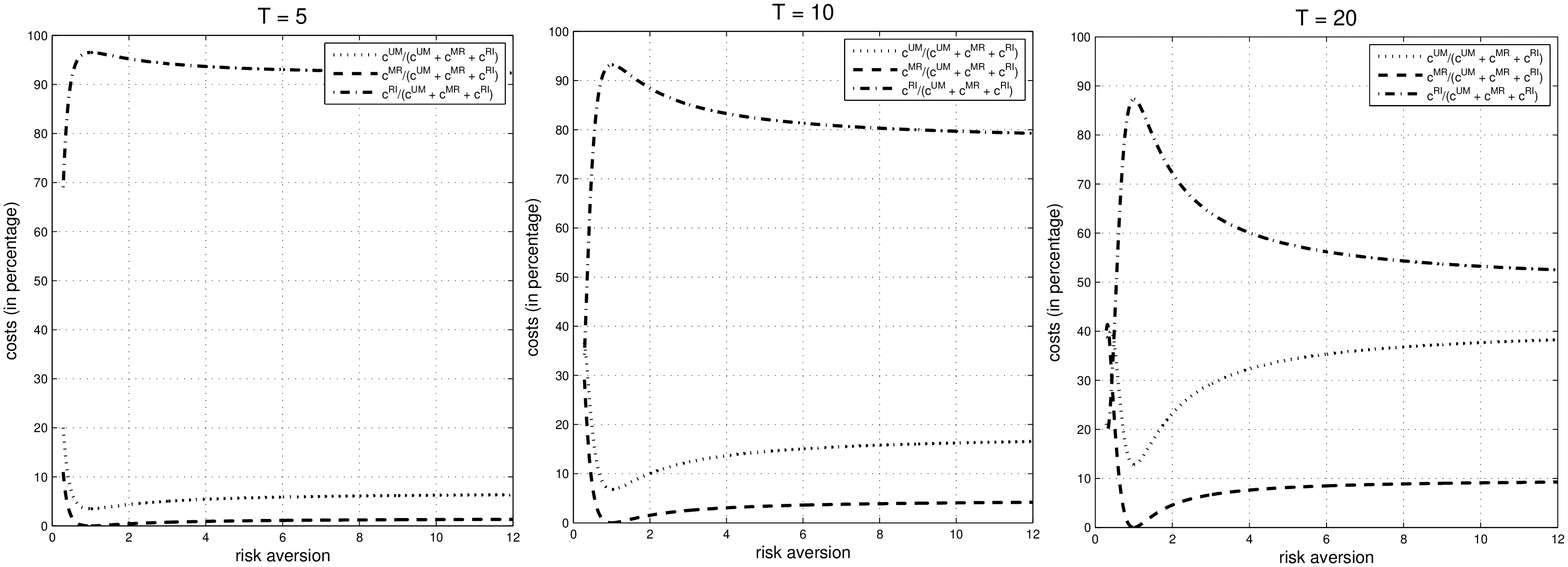}
 \label{figura_costi_rispetto_gamma}
\end{figure}

To gain insights about the importance of predictability, learning, and full information in portfolio management, we assume that time is measured in years and plot the above costs: (i) against investment horizon $T$, for $\gamma=0.8, 1,3$, in Figures \ref{figura_costi_rispetto_T} and \ref{figura_costi_rispetto_T_lungo}, and (ii)  against risk aversion $\gamma$, for $T=5,10,20$, in Figure \ref{figura_costi_rispetto_gamma}. In particular, for each value of $\gamma$ (in Figures \ref{figura_costi_rispetto_T} and \ref{figura_costi_rispetto_T_lungo}) and $T$ (in Figure \ref{figura_costi_rispetto_gamma}), we provide two graphs, vertically aligned: the upper graph represents $C^{UM}$, $C^{MR}$ and $C^{RI}$; the lower graph shows the relative contribution (in percentage) of each cost.
We plot all costs for investment horizons $T$ not exceeding $30$ and $\gamma \leq 12$,
whereas, for the other parameters we use a specification estimated in Brennan \cite{brennan98}.
%\footnote{Our work completes the analysis in Brennan \cite{brennan98} with the study of the effects of learning on the investor's expected utility.}
In particular, the risk free rate is $5\%$, and the annual standard deviation $\sigma$ of the risky asset on the market is $20.2\%$ (see Brennan \cite{brennan98}, Section 3, for a detailed explanation about the derivation of these values). On what concerns the prior of $\Theta$, given a value for $\sigma$, we consider, again as in Brennan \cite{brennan98}, $\Theta$ normally distributed with mean $\theta_0 = 0.08/\sigma$ and variance $v_0=(0.0243/\sigma)^2$.

Table \ref{table-0243} reports the numerical values of the cost of predictability $C^{UM}$, the cost of learning $C^{MR}$, the cost of full information $C^{RI}$, the cost of the three previous features all together $C^{UI}$ and the absolute value of the error $E$ in decomposition (\ref{cost-approximation-analitically}) (all expressed in \%) for values of $T$ and $\gamma$ commonly used in literature for these simulations.\footnote{Cf. Brennan \cite{brennan98}, Xia \cite{xia01}, Honda \cite{honda-03}, Cvitani\'{c} et al. \cite{cvitanic-lazrak-martellini-zapatero-06}, Larsen and Munk \cite{larsen-munk-2012} and Longo and Mainini \cite{longo-mainini-2016}.} The computations are performed, as in Brennan \cite{brennan98}, for two specifications of the annual standard deviation of the risky asset on the market $\sigma$: that is, $20.2\%$ and $14\%$. As for figures, $\Theta$ is normally distributed with mean $0.08/\sigma$ and variance $v_0=(0.0243/\sigma)^2$.\footnote{In Brennan \cite{brennan98} it is also considered $\Theta$ normally distributed with mean $0.08/\sigma$ and variance $v_0=(0.0452/\sigma)^2$. As results are qualitatively similar, we do not report the details.}

\begin{table}
\caption{Cumulated costs for $\theta_0 = 0.08/\sigma$ and $v_0 = (0.0243/\sigma)^2$.}\centering
 \label{table-0243}
\scalebox{0.7}{
\begin{tabular}{cccccccccccc}
\toprule
 & $T$ & \multicolumn{5}{c}{$\sigma=0.202$} & \multicolumn{5}{c}{$\sigma=0.140$}\\
\cmidrule(l){3-7}\cmidrule(l){8-12}
 &  &$C^{UM}$ & $C^{MR}$ & $C^{RI}$ & $C^{UI}$ & $\vert E \vert$ & $C^{UM}$ & $C^{MR}$ & $C^{RI}$ & $C^{UI}$ & $\vert E \vert$ \\
 &  & ($\%$) & ($\%$) &  ($\%$) & ($\%$) & ($\%$) & ($\%$) &  ($\%$) & ($\%$) & ($\%$) & ($\%$)\\
\midrule
\multirow{4}*{$\gamma=11$}
& 30  &2.17 &0.46 &1.39 &3.97 &0.05   &14.23&1.93 &2.18 &17.72 &0.62\\
& 20  &0.74 &0.18 &1.03 &1.93 &0.02   &5.15 &0.91 &1.72 &7.63  &0.15\\
& 10  &0.12 &0.03 &0.58 &0.72 &0.01   &0.82 &0.20 &1.06 &2.06  &0.02 \\
& 5	  &0.02 &0.00 &0.31 &0.33 &0.00   &0.13 &0.03 &0.60 &0.76  &0.00 \\
& 1	  &0.00 &0.00 &0.06 &0.07 &0.01  &0.00 &0.00 &0.13 &0.14  &0.01 \\
\midrule
\multirow{4}*{$\gamma=6$}
& 30  &3.46 &0.74 &2.56 &6.63 &0.13   &21.83&3.23 &4.04 &27.41 &1.69 \\
& 20  &1.19 &0.28 &1.89 &3.33 &0.03   &8.11 &1.50 &3.18 &12.37 &0.42\\
& 10  &0.20 &0.05 &1.06 &1.30 &0.01   &1.32 &0.31 &1.95 &3.55  &0.03 \\
& 5	  &0.04 &0.01 &0.56 &0.61 &0.00   &0.22 &0.05 &1.10 &1.36  &0.01 \\
& 1	  &0.00 &0.00 &0.12 &0.12 &0.00   &0.00 &0.00 &0.24 &0.25  &0.01 \\
\midrule
\multirow{4}*{$\gamma=4$}
& 30  &4.41 &0.96 &3.86 &8.97 &0.26   &26.97 &4.31 &6.12 &34.40 &3.00\\
& 20  &1.53 &0.36 &2.85 &4.68 &0.06   &10.22 &1.96 &4.82 &16.22 &0.78\\
& 10  &0.26 &0.06 &1.59 &1.91 &0.00   &1.70  &0.40 &2.94 &4.97  &0.07\\
& 5	  &0.05 &0.01 &0.85 &0.90 &0.01   &0.29  &0.07 &1.65 &2.00  &0.01\\
& 1	  &0.00 &0.00 &0.18 &0.18 &0.00   &0.01  &0.00 &0.37 &0.37  &0.01\\
\midrule
\multirow{4}*{$\gamma=3$}
& 30  &4.93 &1.07 &5.18 &10.82&0.36   &29.54 &5.03 &8.25 &38.60 &4.22\\
& 20  &1.74 &0.40 &3.81 &5.86 &0.09   &11.32 &2.23 &6.48 &18.92 &1.11\\
& 10  &0.31 &0.06 &2.13 &2.49 &0.01   &1.93  &0.44 &3.94 &6.20  &0.11\\
& 5	  &0.06 &0.01 &1.13 &1.20 &0.00   &0.34  &0.07 &2.21 &2.61  &0.01\\
& 1	  &0.00 &0.00 &0.24 &0.24 &0.00   &0.01  &0.00 &0.49 &0.50  &0.00\\
\midrule
\multirow{4}*{$\gamma=1$}
& 30  &3.61 &0.00 &16.50 &19.51  &0.60 &12.19   &0.00   &27.53 &36.36 &3.36\\
& 20  &1.75 &0.00 &11.94 &13.47  &0.22 &6.34	&0.00	&21.01 &26.01 &1.34\\
& 10  &0.48 &0.00 &6.53  &6.98   &0.03 &1.88	&0.00	&12.34 &13.98 &0.24\\
& 5	  &0.12 &0.00 &3.43  &3.55   &0.00 &0.51	&0.00	&6.78  &7.26  &0.03\\
& 1	  &0.01 &0.00 &0.72  &0.72   &0.01 &0.02	&0.00	&1.47  &1.50  &0.01\\
\midrule
\multirow{4}*{$\gamma=0.8$}
& 30  &7.35 &1.20 &21.16 &27.83 &1.88   &37.66   &11.44 &36.24 &64.80 &20.54\\
& 20  &3.02 &0.35 &15.18 &18.03 &0.52   &15.34   &3.27  &27.20 &40.38 &5.43\\
& 10  &0.70 &0.04 &8.24  &8.92  &0.06   &3.30	 &0.39  &15.70 &18.80 &0.59\\
& 5	  &0.17 &0.01 &4.31  &4.48  &0.01   &0.76	 &0.05 &8.55  &9.29  &0.07\\
& 1	  &0.01 &0.00 &0.90  &0.90  &0.01   &0.03	 &0.00 &1.84  &1.87  &0.00\\
\bottomrule
\end{tabular}
}
\end{table}

The following observations are in order:
(a) The upper graphs of Figures \ref{figura_costi_rispetto_T} and \ref{figura_costi_rispetto_T_lungo} show, consistent with the theory, that all costs increase in the investment horizon $T$
although with a lower intensity the higher the risk aversion. It is also evident (see also Table \ref{table-0243}) that costs are moderate in the short run ($T \leq 5$) and, according to the theoretical analysis, almost entirely attributable to full information.
Interesting to notice is the relative contribution of each single cost. One can observe from the lower graphs of Figure \ref{figura_costi_rispetto_T} that, at least for reasonable investment horizons ($T\leq 30$),\footnote{The lower graphs of Figure \ref{figura_costi_rispetto_T_lungo} show that, in some cases, the relative contribution of costs may appear strange for implausible long investment horizons.} the contribution of learning is the less valuable being, on average, less than $10\%$ of the entire cost, whereas, the remaining $90\%$ is divided, in proportions that depend on risk aversion and investment horizon, between predictability and full information.
Moreover, the relative importance of predictability increases in the investment horizon for any given $\gamma$, whereas the contribution of full information goes in the opposite direction. This is reasonable since the Bayesian estimator
(\ref{conditional-mean-2}) converges to $\Theta$ as $t$ increases to infinity. In other words, as new information comes to light, the approximation of $\Theta$ gets better and, consequently, the additional value of full information diminishes.

(b) According to the previous analysis, the upper graphs of Figure \ref{figura_costi_rispetto_gamma} show that costs shrinks to $0$ as risk aversion increases, though not all in a monotonic way.
Furthermore, costs may be relevant for investors with risk aversion less than $1$ also for relatively short investment horizons. For example (see Table \ref{table-0243}), an investor with $\gamma =0.8$, an investment horizon of $20$ years, and $\sigma = 0.140$ may value predictability $15.34\%$, learning $3.27\%$, and full information $27.20\%$ of the initial wealth. Observe that when costs are relevant, such as in the previous specification,  approximation (\ref{cost-approximation}) is less accurate, as the error may be as high as $20\%$.
Moreover, the lower graphs of Figure \ref{figura_costi_rispetto_gamma} show that, for any given $T$, the relative contribution of predictability and learning increase for $\gamma > 1$ and decrease for $\gamma < 1$, whereas the relative contribution of full information goes in the opposite direction, and this behavior is exacerbated for longer investment horizons.
This means that the logarithmic investor (\textit{i.e.}, $\gamma =1$) not only gives zero value to learning (\textit{i.e.}, $C^{MR}=0$), which is well known in literature (cf. Kuwana \cite{kuwana-95}), but also assigns the lowest relative value to predictability (although this value increases in the investment horizon as the middle-lower graphs of Figures \ref{figura_costi_rispetto_T} and \ref{figura_costi_rispetto_T_lungo} show). In other words, the logarithmic investor gives, among CRRA investors, the lowest relative value to conditional information.

(c) As already noticed, the welfare effect of learning
is the less valuable in the context of dynamic portfolio choice with parameter uncertainty. This is consistent with the findings of Haugh et al. \cite{haugh-kogan-wang-2006} and Larsen and Munk \cite{larsen-munk-2012}. However, Haugh et al. \cite{haugh-kogan-wang-2006} analysis leaves room for the  possibility that predictability may have relevant welfare effects and, indeed, for reasonable parameters' values, our simulations show that this is the case for sufficiently long investment horizons.
%and/or higher risk aversion.
It is worth noticing that, despite the fact that the cost of behaving myopically is not very relevant,
the hedging demand, defined as the difference  $\pi_{\gamma}^{R}-\pi_{\gamma}^{M}$ (see (\ref{optimal-portfolio-learner-power}) and (\ref{optimal-portfolio-myopic-power})), associated to learning may be substantial as it is shown in Brennan \cite{brennan98} and, in a more general setting, in Longo and Mainini \cite{longo-mainini-2016} (see also Haugh et al. \cite{haugh-kogan-wang-2006}).

\section{DIY \textit{vs} Delegated portfolio management}\label{Section-DIY-investor}

In this section, we use the previous analysis to investigate the saver's decision of whether to manage her/his portfolio personally, \textit{DIY investor}, or hire, at some cost, a professional investor.
We abstract from possible conflicts of interests arising from agency contracts by assuming that financial intermediaries may be compelled by authorities to act according to client's risk profile and financial situation.
\footnote{For example, within the European Economic Community, the \textit{Markets in Financial Instruments Directive 2004/39/CE} (known as \textquotedblleft MiFID\textquotedblright), issued by the European Parliament, on Article 19, paragraph 4, states: \textquotedblleft When providing investment advice or portfolio management the investment firm shall obtain the necessary information regarding the client's or potential client's knowledge and experience in the investment field relevant to the specific type of product or service, his financial situation and his investment objectives so as to enable the firm to recommend to the client or potential client the investment services and financial instruments that are suitable for him\textquotedblright. See also Directive 2006/73/CE articles 35 and 36.}
We consider the case where a saver of type $i \in \left\{ I,R,M,U\right\}$ faces the possibility of delegating, against the payment of a fee, her/his portfolio management to a professional investor of type $j \in \left\{ I,R,M,U\right\}$, with $i \precsim j$.
Since portfolios' management fees are typically on annual basis and expressed as a percentage of the invested wealth, we measure time in years and assume that a saver of type $i$ hires a professional investor of type $j$ as long as the annual commission fee does not exceed the quantity $c^{ij}$, which is the \textit{annual cost} defined by the indifference relation:
\begin{equation}
V^{i}(x;\gamma, T) = V^{j}(x(1-c^{ij})^T;\gamma, T).
\end{equation}
That is,
\begin{eqnarray*}
c^{ij}  = c^{ij}(\gamma, T) & =  &  1-\exp \left(
\frac{ \varphi ^{i}\left( \gamma ,T\right) -\varphi ^{j}\left( \gamma ,T\right) }{(1-\gamma)T }\theta _{0}^{2} +
\frac{\psi ^{i}\left( \gamma ,T\right) -\psi ^{j}\left( \gamma,T\right) }{(1-\gamma)T }
\right) \\
 & =  &  1-(1-C^{ij})^{1/T},
\end{eqnarray*}
where $C^{ij}$ is the cumulated cost defined in (\ref{cumulated-costs-definition}). Notice that, costs are again independent of the initial wealth $x$ and logarithmic costs (\textit{i.e.}, $\gamma=1$),
\begin{eqnarray}
& &c^{UM}\left( 1,T\right)  =1-e^{-v_{0}/2}\left( 1+v_{0}T\right) ^{1/(2T)}, \label{c-log-UM} \\
& &c^{MR}\left( 1,T\right)  =0, \label{c-log-MR} \\
& &c^{RI}\left( 1,T\right)  =1-\left( 1+v_{0}T\right) ^{-1/(2T)}, \label{c-log-RI}
\end{eqnarray}
are equal to the limits, as $\gamma \rightarrow 1$, of the corresponding power costs.
$c^{ij}$ is an annual version of the cost (\ref{cumulated-costs-definition}) (for an explanation of the definition, we refer the reader to  Brennan and Torous \cite{brennan-torous-1999}, Section 2.3) and may be seen as the maximum annual management fee, in terms of fraction of initial wealth, that a saver of type $i$, risk aversion $\gamma$ and investment horizon $T$, is prepared to pay for hiring a professional investor of type $j$, with $i\precsim j$. Moreover, a decomposition similar to (\ref{cost-approximation}) holds also in this case, that is
\begin{equation}
 c^{UI}\approx c^{UM}+c^{MR}+ c^{RI},   \label{cost-approximation-annual}
\end{equation}
and it can be used to assess the importance of predictability, learning and full information in the saver decision of delegating the portfolio management: it tells us about the more valuable
investment advisory services by a saver.
In fact, $c^{UM}$, $c^{MR}$ and $c^{RI}$ may be considered the parts of a management fee attributable, respectively, to advices about predictability, learning and full information.

\begin{figure}
\caption{Annual costs against investment horizon $T \,(\leq 30)$ for $\sigma=0.202$, $\theta_0 = 0.08/\sigma$ and $v_0=(0.0243/\sigma)^2$.}
\includegraphics[width=1\columnwidth]{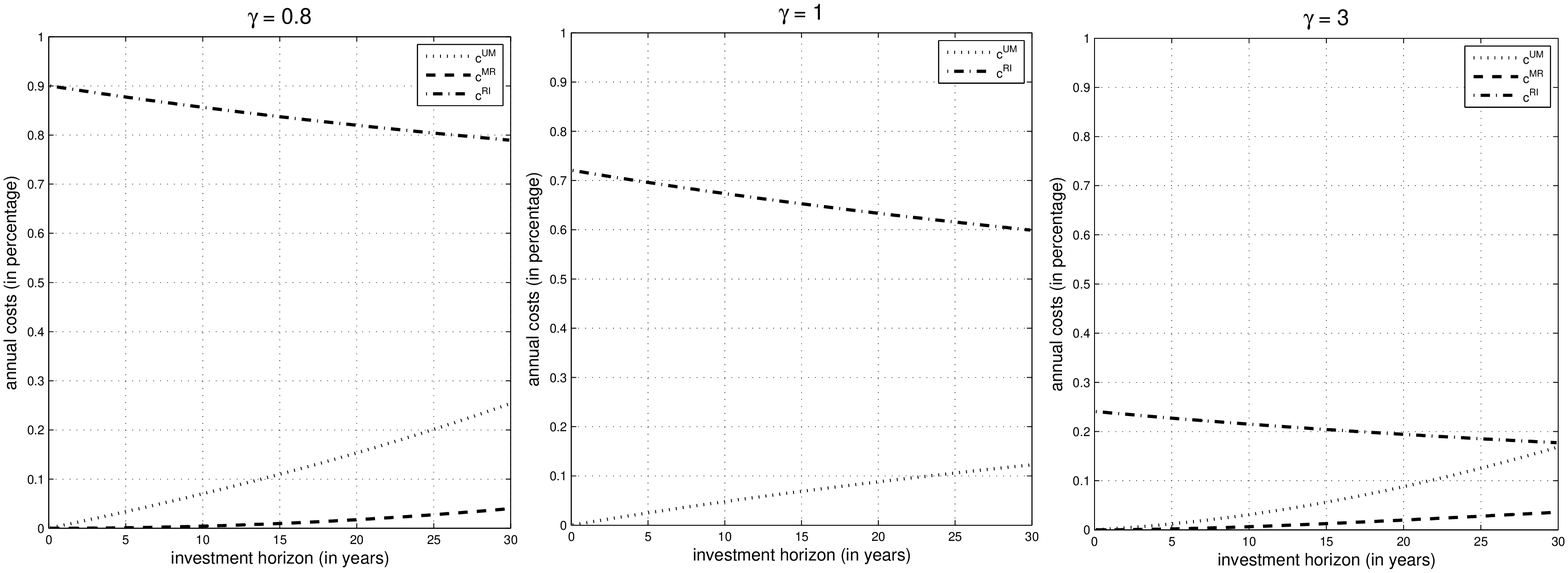} \\
\includegraphics[width=1\columnwidth]{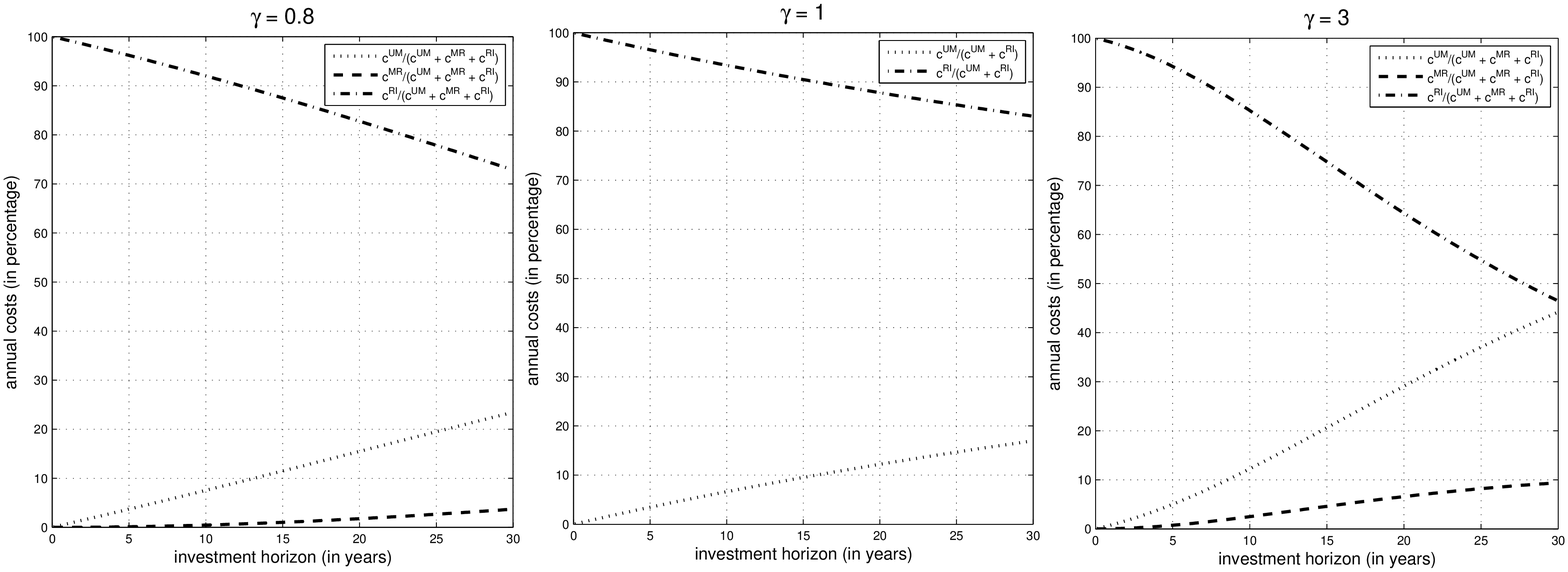}
 \label{figura_costi_annui_rispetto_T}
\end{figure}

\begin{figure}
\caption{Annual costs against investment horizon $T \, (\leq 250)$ for $\sigma=0.202$, $\theta_0 = 0.08/\sigma$ and $v_0=(0.0243/\sigma)^2$.}
\includegraphics[width=1\columnwidth]{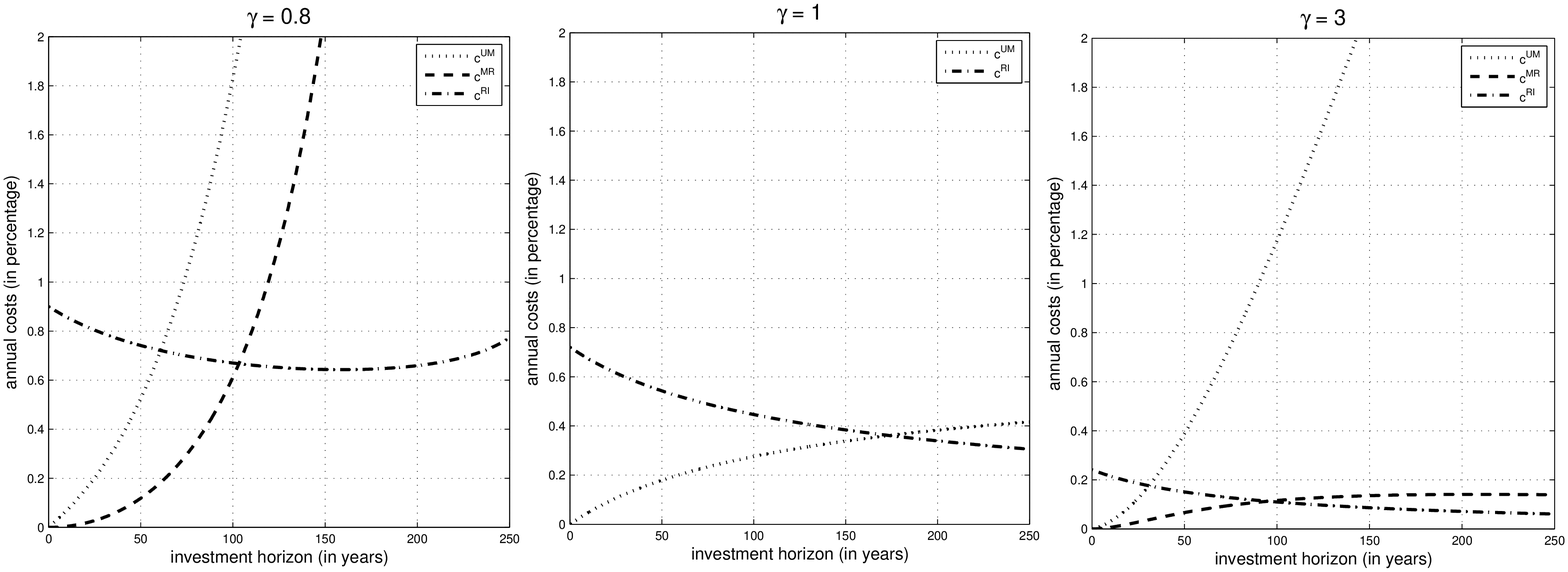} \\
\includegraphics[width=1\columnwidth]{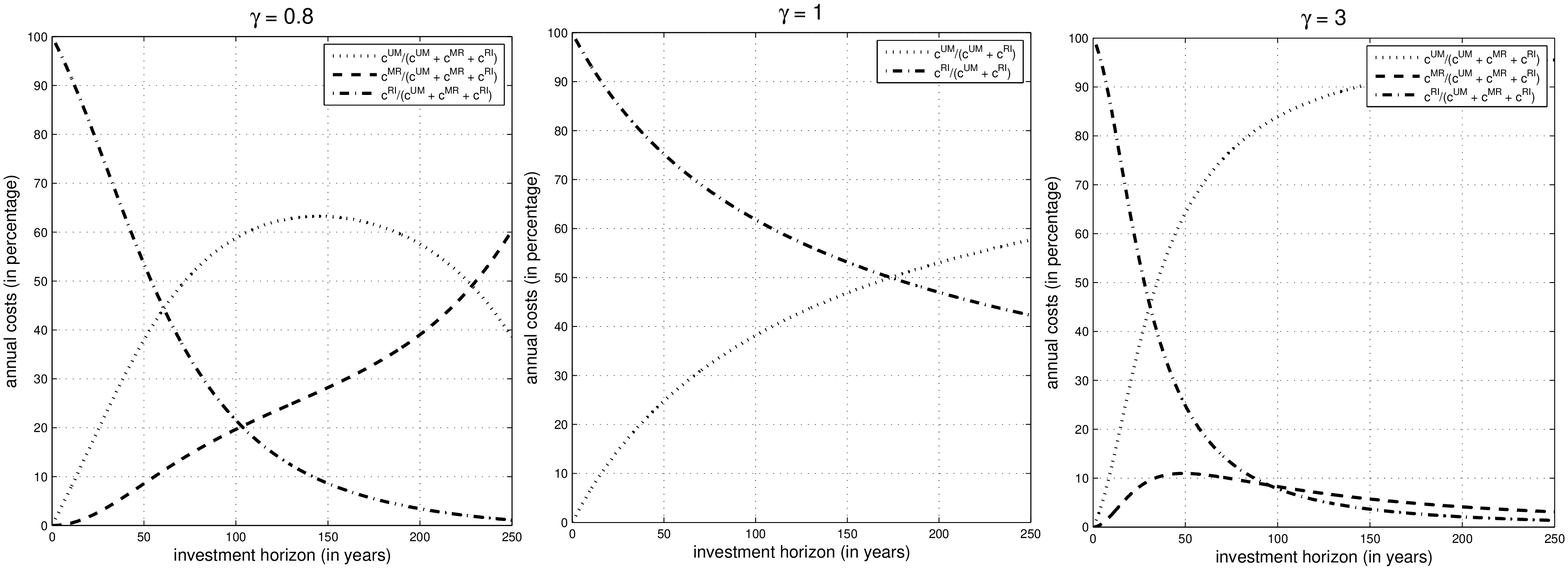}
 \label{figura_costi_annui_rispetto_T_lungo}
\end{figure}

\begin{figure}
\caption{Annual costs against risk aversion $\gamma \, (\leq 12)$ for $\sigma=0.202$, $\theta_0 = 0.08/\sigma$ and $v_0=(0.0243/\sigma)^2$.}
\includegraphics[width=1\columnwidth]{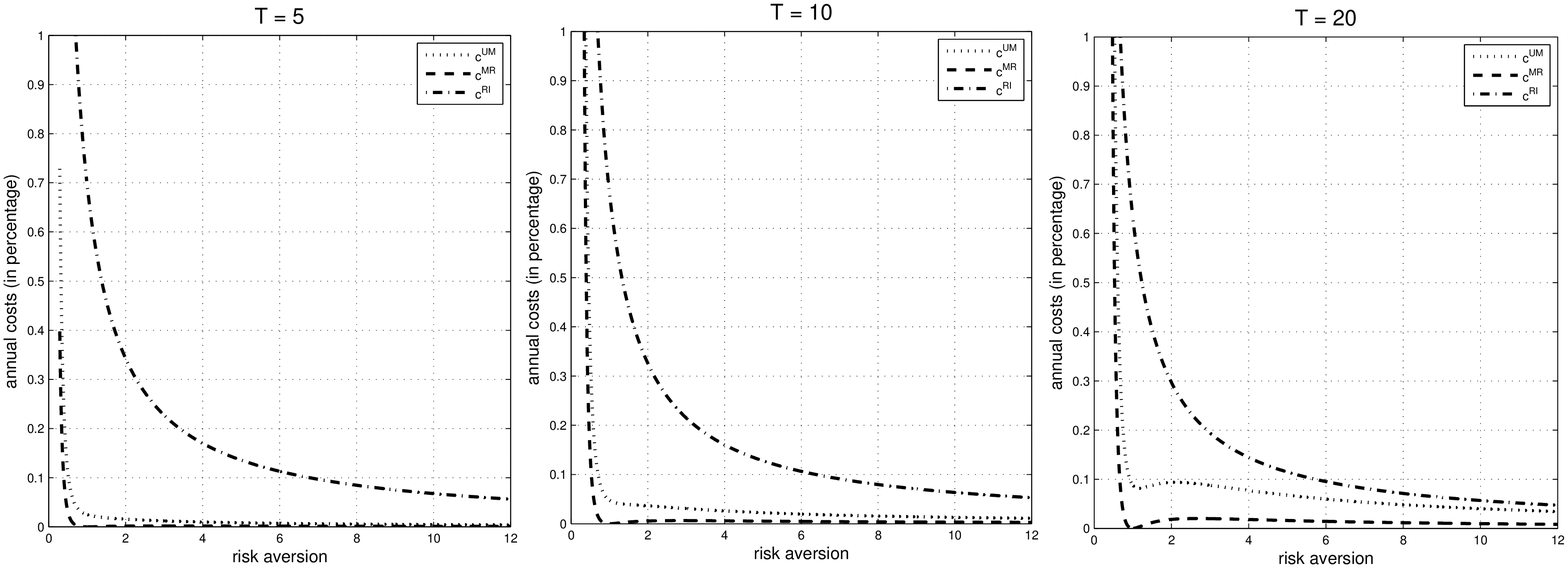} \\
\includegraphics[width=1\columnwidth]{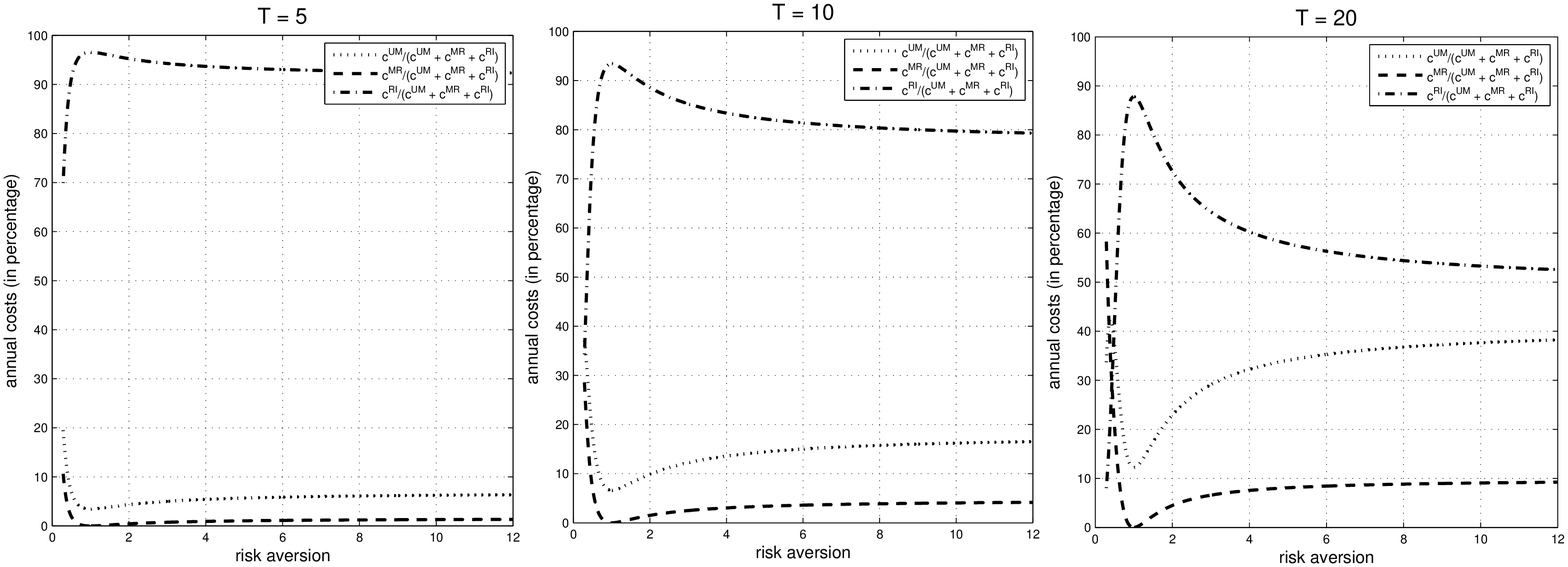}
 \label{figura_costi_annui_rispetto_gamma}
\end{figure}

Similarly to the previous section, we first analyse costs $ c^{UM}$, $c^{MR}$, $c^{RI}$ and
approximation (\ref{cost-approximation-annual}) in generality with respect to risk aversion and investment horizon and then simulations analogous to those of the previous section will provide practical insights.
Indeed, Table \ref{table-0243-annual}, whose description is similar to Table \ref{table-0243}, reports the annual costs for the same parameter configuration of Table \ref{table-0243}. Figures \ref{figura_costi_annui_rispetto_T}, \ref{figura_costi_annui_rispetto_T_lungo} and \ref{figura_costi_annui_rispetto_gamma} represent respectively the annual versions of Figures \ref{figura_costi_rispetto_T}, \ref{figura_costi_rispetto_T_lungo} and \ref{figura_costi_rispetto_gamma}.

We start from the investment horizon $T$. For each $\gamma > 0$, the following Taylor expansions hold in a (right) neighborhood of $T=0$:
\begin{eqnarray}
& &c^{UM}\left( \gamma ,T\right)  = \frac{v_{0}^{2}}{4\gamma }T+o\left( T\right), \\
& &c^{MR}\left( \gamma ,T\right)  = o\left(T\right) ,\\
& &c^{RI}\left( \gamma ,T\right)  =  1-\exp \left( -\frac{v_{0}}{2\gamma }\right) +\exp \left( -\frac{v_{0}}{2\gamma }\right) \frac{v_{0}^{2}}{2\gamma^{2}}\left( \frac{1}{2}-\gamma \right) T+o\left( T\right), \\
& &c^{UI}\left( \gamma ,T\right)  = 1-\exp \left( -\frac{v_{0}}{2\gamma }\right) +\exp \left( -\frac{v_{0}}{
2\gamma }\right) \frac{v_{0}^{2}}{2\gamma ^{2}}\left( \frac{1}{2}-\frac{
\gamma }{2}\right) T+o\left( T\right) .
\end{eqnarray}
Hence, contrary to the cumulated case, not all costs vanish as $T$ goes to $0$: indeed,
\begin{equation}
\lim_{T \rightarrow 0 }~c^{UM}\left( \gamma ,T\right) = \lim_{T \rightarrow 0 }~c^{MR}\left( \gamma ,T\right) =0
\end{equation}
but
\begin{equation}
\lim_{T \rightarrow  0}~c^{RI}\left( \gamma ,T\right) = 1-\exp \left( -\frac{v_{0}}{2\gamma }\right)>0.
\end{equation}
Consistent with the previous section, in the short run, also annual costs are mainly explained by full information, then by predictability and residually by learning. We can say even more, for very short investment horizons delegation is primarily, if not exclusively, motivated by the beliefs that professional investment advisers are better informed (\textquotedblleft insiders\textquotedblright) rather than their ability of gathering and processing information or learning from financial data, and this is reasonable since within this time lag there is no room for learning nor for processing information.
Moreover,
\begin{equation}  \label{cost-approximation-analitically-annual}
 c^{UI}\left( \gamma,T\right)= c^{UM}\left(\gamma, T\right) +c^{MR}\left(\gamma, T\right)+ c^{RI}\left( T\right) + e\left(\gamma, T\right),
\end{equation}
where the \textit{error} $e(\gamma, T)$ is such that
\begin{equation}  \label{cost-approximation-analitically-annual-error}
e\left(\gamma, T\right) =  \left( \exp \left( -\frac{v_{0}}{2\gamma } \right) -1 \right) \frac{v_{0}^{2}}{4\gamma }T+o\left( T\right),\;\;\; T\rightarrow 0.
\end{equation}
Despite the fact that approximation (\ref{cost-approximation-annual}) is less accurate than (\ref{cost-approximation}) (compare the errors (\ref{cost-approximation-analitically-error}) and (\ref{cost-approximation-analitically-annual-error})), it is still extremely precise, as simulations ahead show (see error columns in Table \ref{table-0243-annual}), as long as the ratio $v_0/\gamma$ is sufficiently close to $0$, and this is true for financial significant ranges of parameters $v_0$ and $\gamma$, and financial reasonable investment horizons (for instance, $T\leq 30$).

Also in this case, costs' behavior as the investment horizon increases to infinity depends on $\gamma$. Again, we distinguish three cases: \textit{(i)} $\gamma > 1$, costs exist for all $T > 0$ and \begin{equation}
\lim_{T\rightarrow \infty }~c^{UM}\left( \gamma ,T\right) = 1, \;\;\;
\lim_{T\rightarrow \infty }~c^{MR}\left( \gamma ,T\right) =\lim_{T\rightarrow \infty }~c^{RI}\left( \gamma ,T\right) =0;
\end{equation}
\textit{(ii)} $\gamma = 1$ (the logarithmic case, see (\ref{c-log-UM})-(\ref{c-log-RI})), costs exist for all $T> 0$, $c^{MR}\left( 1 ,T\right)=0$, for all $T$, and
\begin{equation}
\lim_{T\rightarrow \infty }~c^{UM}\left( 1 ,T\right) = 1-\exp \left( -\frac{v_{0}}{2} \right), \;\;\; \lim_{T\rightarrow \infty }~c^{RI}\left( 1 ,T\right) =0;
\end{equation}
\textit{(iii)} $0 < \gamma <1$, costs exist for all $T < \bar{T}$ and
\begin{equation}
\lim_{T\rightarrow \bar{T}}~c^{UM}\left( \gamma ,T\right) = c^{UM}\left( \gamma ,\bar{T}\right)<1, \;\;% \; \textcolor{red}{\textbf{(vero?)}},
\lim_{T\rightarrow \bar{T}}~c^{MR}\left( \gamma ,T\right) = % 1 \; \textcolor{red}{\textbf{(vero?)}}, \; \;
\lim_{T\rightarrow \bar{T}}~c^{RI}\left( \gamma ,T\right) = 1.
\end{equation}
An important observation concerns costs monotonicity with respect to $T$. We consider the more significant cases of $\gamma \geq 1$. For $\gamma >1$, contrary to the cumulated versions, none of the annual costs is increasing for all $T$. In particular: \textit{(i)} $c^{UM}$ is increasing in a right neighborhood of $0$  and definitely but may fail to be increasing in some middle interval for $\theta_0$ sufficiently small; \textit{(ii)} $c^{MR}$ is increasing in a right neighborhood of $0$ and decreasing definitely; \textit{(iii)} $c^{RI}$ is decreasing for all $T>0$. The latter behavior is reasonable as benefits from full information (we recall that $c^{RI}$ is a measure of value of full information) decrease in the investment horizon since the longer the observation interval, the better the approximation of unobservable quantities. For $\gamma =1$ (\textit{i.e.}, the logarithmic case, see (\ref{c-log-UM})-(\ref{c-log-RI})), it can be proved that, for all $T>0$, $c^{UM}$ is increasing and $c^{RI}$ decreasing.

We now turn to the analysis with respect to the parameter $\gamma$. For any fixed $T>0$, costs exist for all $\gamma > \bar{\gamma}$ and we have
\begin{equation}
\lim_{\gamma \rightarrow \infty }~c^{UM}\left( \gamma ,T\right) = \lim_{\gamma \rightarrow \infty }~c^{MR}\left( \gamma ,T\right) =
\lim_{\gamma \rightarrow \infty }~c^{RI}\left( \gamma ,T\right) = 0.
\end{equation}
Again, for $\gamma$ sufficiently high, costs become very small and approximation (\ref{cost-approximation-annual}) is accurate. The explanation for these behaviors is similar to the cumulated version.
As already noticed, costs tend to the logarithmic case as $\gamma$ approaches to $1$, and, as $\gamma \rightarrow \bar{\gamma}$, we have
\begin{equation}
\lim_{\gamma \rightarrow \bar{\gamma}}~c^{UM}\left( \gamma ,T\right) = c^{UM}\left( \bar{\gamma} ,T\right)<1, \;
\lim_{\gamma\rightarrow \bar{\gamma}}~c^{MR}\left( \gamma ,T\right) =
\lim_{\gamma\rightarrow \bar{\gamma}}~c^{RI}\left( \gamma ,T\right) = 1.
\end{equation}
Finally, similar to the cumulated cases, $c^{RI}(\gamma, T)$ is decreasing with respect to $\gamma$ for all $T>0$, whereas simulations ahead show that other costs are not monotonic in general.

\begin{table}
\caption{Annual costs for $\theta_0 = 0.08/\sigma$ and $v_0 = (0.0243/\sigma)^2$.}\centering
 \label{table-0243-annual}
\scalebox{.7}{
\begin{tabular}{cccccccccccc}
\toprule
 & $T$ & \multicolumn{5}{c}{$\sigma=0.202$} & \multicolumn{5}{c}{$\sigma=0.140$}\\
\cmidrule(l){3-7}\cmidrule(l){8-12}
 &  &$c^{UM}$ & $c^{MR}$ & $c^{RI}$ & $c^{UI}$ & $\vert e \vert$ & $c^{UM}$ & $c^{MR}$ & $c^{RI}$ & $c^{UI}$ & $\vert e \vert$ \\
 &  & ($\%$) & ($\%$) &  ($\%$) & ($\%$) & ($\%$) & ($\%$) &  ($\%$) & ($\%$) & ($\%$) & ($\%$)\\
\midrule
\multirow{4}*{$\gamma=11$}
& 30  &0.07 &0.02 &0.05 &0.13 &0.01   &0.51 &0.06 &0.07 &0.65  &0.01\\
& 20  &0.04 &0.01 &0.05 &0.10 &0.00   &0.26 &0.05 &0.09 &0.40  &0.00\\
& 10  &0.01 &0.00 &0.06 &0.07 &0.00   &0.08 &0.02 &0.11 &0.21  &0.00 \\
& 5	  &0.00 &0.00 &0.06 &0.07 &0.01  &0.03 &0.01 &0.12 &0.15  &0.01 \\
& 1	  &0.00 &0.00 &0.06 &0.07 &0.01  &0.00 &0.00 &0.13 &0.14  &0.01 \\
\midrule
\multirow{4}*{$\gamma=6$}
& 30  &0.12 &0.02 &0.09 &0.23 &0.00   &0.82 &0.11 &0.14 &1.06 &0.01 \\
& 20  &0.06 &0.01 &0.10 &0.17 &0.00   &0.42 &0.08 &0.16 &0.66 &0.00\\
& 10  &0.02 &0.00 &0.11 &0.13 &0.00   &0.13 &0.03 &0.20 &0.36  &0.00 \\
& 5	  &0.01 &0.00 &0.11 &0.12 &0.00   &0.04 &0.01 &0.22 &0.27  &0.00 \\
& 1	  &0.00 &0.00 &0.12 &0.12 &0.00   &0.00 &0.00 &0.24 &0.25  &0.01 \\
\midrule
\multirow{4}*{$\gamma=4$}
& 30  &0.15 &0.03 &0.13 &0.31 &0.00   &1.04  &0.15 &0.21 &1.40  &0.00\\
& 20  &0.08 &0.02 &0.14 &0.24 &0.00   &0.54  &0.10 &0.25 &0.88  &0.01\\
& 10  &0.03 &0.01 &0.16 &0.19 &0.01   &0.17  &0.04 &0.30 &0.51  &0.00\\
& 5	  &0.01 &0.00 &0.17 &0.18 &0.00   &0.06  &0.01 &0.33 &0.40  &0.00\\
& 1	  &0.00 &0.00 &0.18 &0.18 &0.00   &0.01  &0.00 &0.37 &0.37  &0.01\\
\midrule
\multirow{4}*{$\gamma=3$}
& 30  &0.17 &0.04 &0.18 &0.38 &0.01   &1.16  &0.17 &0.29 &1.61 &0.01\\
& 20  &0.09 &0.02 &0.19 &0.30 &0.00   &0.60  &0.11 &0.33 &1.04 &0.00\\
& 10  &0.03 &0.01 &0.22 &0.25 &0.01   &0.19  &0.04 &0.40 &0.64 &0.01\\
& 5	  &0.01 &0.00 &0.23 &0.24 &0.00   &0.07  &0.01 &0.45 &0.53 &0.00\\
& 1	  &0.00 &0.00 &0.24 &0.24 &0.00   &0.01  &0.00 &0.49 &0.50 &0.00\\
\midrule
\multirow{4}*{$\gamma=1$}
& 30  &0.12 &0.00 &0.60  &0.72   &0.00    &0.43 &0.00   &1.07  &1.50  &0.00\\
& 20  &0.09 &0.00 &0.63  &0.72   &0.00    &0.33	&0.00	&1.17  &1.50  &0.00\\
& 10  &0.05 &0.00 &0.67  &0.72   &0.00    &0.19	&0.00	&1.31  &1.50  &0.00\\
& 5	  &0.02 &0.00 &0.70  &0.72   &0.00    &0.10	&0.00	&1.39  &1.50  &0.01\\
& 1	  &0.01 &0.00 &0.72  &0.72   &0.01    &0.02	&0.00	&1.47  &1.50  &0.01\\
\midrule
\multirow{4}*{$\gamma=0.8$}
& 30  &0.25 &0.04 &0.79  &1.08  &0.00       &1.56  &0.40  &1.49  &3.42 &0.03\\
& 20  &0.15 &0.02 &0.82  &0.99  &0.00       &0.83  &0.17  &1.57  &2.55 &0.02\\
& 10  &0.07 &0.00 &0.86  &0.93  &0.00       &0.33  &0.04  &1.69  &2.06 &0.00\\
& 5	  &0.03 &0.00 &0.88  &0.91  &0.00       &0.15  &0.01  &1.77  &1.93 &0.00\\
& 1	  &0.01 &0.00 &0.90  &0.90  &0.00       &0.03  &0.00  &1.84  &1.87 &0.00\\
\bottomrule
\end{tabular}
}
\end{table}


Simulations suggest the following observations:
(a) For financial reasonable values of the parameters, simulations indeed show (see Table \ref{table-0243-annual}) that approximation (\ref{cost-approximation-annual}) is very accurate for financially plausible investment horizons, such as those not exceeding $30$ years.

(b) Similar to the cumulated version analyzed in the previous section, annual costs, depending on investment horizon and risk aversion, are mainly justified by predictability and/or full information, being the cost (\textit{i.e.}, the value) of learning almost negligible. Consistent with the analytical findings, importance of full information with respect to predictability and learning for very short investment horizons is even striking if compared with the cumulated version.
Moreover, costs may be relevant for investors with risk aversion less than $1$ (aggressive investors) which, consequently, may be expected to be the more interested in delegation.

(c) A last remark is on costs' magnitude that, in some cases, might appear too low compared to real management fees. Similar to the concept of \textit{pure premium} in actuarial jargon, costs of Table  \ref{table-0243-annual} should be considered as \textit{pure} management fees ascribable to the sole investment advisory services that do not include administrative services and overheads, for instance.% and profits.

\section{Conclusion}\label{Section-Conclusion}

In this paper we evaluate the utility effects -- in terms of additional fraction of initial wealth that makes two investment strategies indifferent -- of two different degrees of bounded rationality for a CRRA investor operating in a continuous-time financial market with parameter uncertainty and compare them between each other and with the effect of uncertainty measured by the additional fraction of initial wealth that makes a rational agent indifferent between a full information and a partial information scenario.
Agents may be boundedly rational in that they either do not learn from fluctuations in the conditional distribution of the unobserved variable as more information come to light (\textit{myopic} investors), or they completely disregard available information provided by prices and its predictable role (\textit{unconditional} investors).
The welfare difference between two different scenarios is quantified in terms of fraction of initial wealth needed to obtain the same expected utility.
We find that, for reasonable values of the parameters, the main effects are produced, depending on risk aversion and investment horizon, by full information and/or predictability, being the effect of learning marginal.

We also investigate the saver's decision of whether to manage her/his portfolio personally or hire, against the payment of a management fee, a professional investor. We find that delegation is mainly motivated, depending on risk aversion and investment horizon, by the beliefs that professional investors are either better informed (\textquotedblleft insiders\textquotedblright) or more capable of gathering and processing information rather than their ability of learning from financial data.
In particular, for very short investment horizons, we find the delegation is primarily, if not exclusively, motivated by the beliefs that professional investors are better informed.
Moreover, it is also suggested that aggressive investors (\textit{i.e.}, $\gamma <1$) are presumably the more interested in delegation.

It would be interesting to extend the analysis to more general settings in terms of utility functions and/or uncertainty structure. This is left for further research.

%\section*{Acknowledgements}
%
%An unnumbered section, e.g. \verb"\section*{Acknowledgement(s)}", should be used for thanks, etc.
%and placed before any Funding or References sections.

%\section*{Funding}
%
%VEDERE .... The second author acknowledges Universit\`{a}  Cattolica del Sacro Cuore for financial support under the project (Assegno di ricerca) "....".

\bibliographystyle{abbrv}
\bibliography{DIY-biblio}

\begin{thebibliography}{10}

\bibitem{barberis-thaler-2003}
N.~Barberis and R.~Thaler.
\newblock A survey of behavioral finance.
\newblock In G.~Constantinides, M.~Harris, and R.~Stulz, editors, {\em Handbook
  of the Economics of Finance}, pages 1051--1121. Elsevier Science B.V., 2003.

\bibitem{bjork-davis-landen-2010}
T.~Bj\"ork, M.~Davis, and C.~Land\'en.
\newblock Optimal investment under partial information.
\newblock {\em Mathematical Methods of Operations Research}, 71:371--399, 2010.

\bibitem{branger-larsen-munk-2013}
N.~Branger, L.~Larsen, and C.~Munk.
\newblock Robust portfolio choice with ambiguity and learning about return
  predictability.
\newblock {\em Journal of Banking and Finance}, 37:1397--1411, 2013.

\bibitem{brendle-06}
S.~Brendle.
\newblock Portfolio selection under incomplete information.
\newblock {\em Stochastic Processes and their Applications}, 106:701--723,
  2006.

\bibitem{brennan98}
M.~Brennan.
\newblock The role of learning in dynamic portfolio decisions.
\newblock {\em European Finance Review}, 1:295--306, 1998.

\bibitem{brennan-torous-1999}
M.~Brennan and W.~Torous.
\newblock Individual decision making and investor welfare.
\newblock {\em Economic Notes}, 28:119--143, 1999.

\bibitem{brennan-xia-2010}
M.~Brennan and Y.~Xia.
\newblock Persistence, predictability, and portfolio planning.
\newblock In C.-F. Lee, A.~Lee, and J.~Lee, editors, {\em Handbook of
  Quantitative Finance and Risk Management}, pages 289--318. Springer, 2010.

\bibitem{browne-whitt-1996}
S.~Browne and W.~Whitt.
\newblock Portfolio choice and the bayesian kelly criterion.
\newblock {\em Advances in Applied Probability}, 28:1145--1176, 1996.

\bibitem{cvitanic-lazrak-martellini-zapatero-06}
J.~Cvitani\'{c}, A.~Lazrak, L.~Martellini, and F.~Zapatero.
\newblock Dynamic portfolio choice with parameter uncertainty and the economic
  value of analysts’ recommendations.
\newblock {\em The Review of Financial Studies}, 19:1113--1156, 2006.

\bibitem{debondt-thaler-95}
W.~De~Bondt and R.~Thaler.
\newblock Financial decision-making in markets and firms: A behavioral
  perspective.
\newblock In R.~Jarrow, V.~Maksimovic, and W.~Ziemba, editors, {\em Handbooks
  in Operations Research and Management Science, Volume 9}, pages 385--410.
  Elsevier, Amsterdam, 1995.

\bibitem{haugh-kogan-wang-2006}
M.~Haugh, L.~Kogan, and J.~Wang.
\newblock Evaluating portfolio policies: A duality approach.
\newblock {\em Operations Research}, 54:405--418, 2006.

\bibitem{honda-03}
T.~Honda.
\newblock Optimal portfolio choice for unobservable and regime-switching mean
  returns.
\newblock {\em Journal of Economic Dynamics and Control}, 28:45--78, 2003.

\bibitem{huang-liu-2007}
L.~Huang and H.~Liu.
\newblock Rational inattention and portfolio selection.
\newblock {\em The Journal of Finance}, 62:1999--2040, 2007.

\bibitem{karatzas-zhao-01}
I.~Karatzas and X.~Zhao.
\newblock Bayesian adaptive portfolio optimization.
\newblock In E.~Jouini, J.~Cvitani\'c, and M.~Musiela, editors, {\em Option
  Pricing, Interest Rates and Risk Management}, pages 632--669. Cambridge
  University Press, 2001.

\bibitem{kuwana-95}
Y.~Kuwana.
\newblock Certainty equivalence and logarithmic utilities in
  consumption/investment problems.
\newblock {\em Mathematical Finance}, 5:297--310, 1995.

\bibitem{lakner-95}
P.~Lakner.
\newblock Utility maximization with partial information.
\newblock {\em Stochastic Processes and their Applications}, 56:247--273, 1995.

\bibitem{larsen-munk-2012}
L.~Larsen and C.~Munk.
\newblock Robust portfolio choice with ambiguity and learning about return
  predictability.
\newblock {\em Journal of Economics Dynamics \& Control}, 36:266--293, 2012.

\bibitem{longo-mainini-2016}
M.~Longo and A.~Mainini.
\newblock Learning and portfolio decisions for \textsc{CRRA} investors.
\newblock {\em International Journal of Theoretical and Applied Finance}, 19,
  1650018, 2016.

\bibitem{pastor-veronesi-nber-2009}
L.~Pastor and P.~Veronesi.
\newblock Learning in financial markets.
\newblock {\em NBER Working Paper 14646}, 2009.

\bibitem{pham-2011}
H.~Pham.
\newblock Portfolio optimization under partial observation: theoretical and
  numerical aspects.
\newblock In D.~Crisan and B.~Rozovsky, editors, {\em The Oxford Handbook of
  Nonlinear Filtering}, pages 990--1018. Oxford University Press, 2011.

\bibitem{rieder-bauerle}
U.~Rieder and N.~B\"auerle.
\newblock Portfolio optimization with unobservable markov-modulated drift
  process.
\newblock {\em Journal of Applied Probability}, 42:362--378, 2005.

\bibitem{rishel99}
R.~Rishel.
\newblock Optimal portfolio management with partial observation and power
  utility function.
\newblock In W.~McEneany, G.~Yin, and Q.~Zhang, editors, {\em Stochastic
  Analysis, Control, Optimization and Applications: Volume in Honor of W. H.
  Fleming}, pages 605--620. Birkh\"auser Verlag, Boston, 1999.

\bibitem{rogers-01}
L.~Rogers.
\newblock The relaxed investor and parameter uncertainty.
\newblock {\em Finance and Stochastics}, 5:131--154, 2001.

\bibitem{xia01}
Y.~Xia.
\newblock Learning about predictability: The effects of parameter uncertainty
  on dynamic asset allocation.
\newblock {\em The Journal of Finance}, 56:205--246, 2001.

\end{thebibliography}

\section{Appendix}
We sketch the main steps in the derivation of indirect utilities (\ref{utilita-indiretta-informed-power}), (\ref{utilita-indiretta-learner-power}), (\ref{utilita-indiretta-myopic-power}) and (\ref{utilita-indiretta-static-power}). We treat the power case (\textit{i.e.}, $\gamma \neq 1$),  the logarithmic case (\textit{i.e.}, $\gamma =1$) can be treated similarly. Moreover, it can be proved that each logarithmic case is the limit of the power case as $\gamma \to 1$ (more precisely, for each $i$, $V^{i}(x;1,T)$ is the limit of $V^{i}(x;\gamma,T)-1/(1-\gamma) $, as $\gamma \to 1$). Recall the definition $\hat{\Theta}(t,y)=\left( \theta _{0}+v_{0}y\right) /\left( 1+v_{0}t\right) $ (see (\ref{conditional-mean})), hence $\hat{\Theta}_{y}=v_{0}/\left( 1+v_{0}t\right) $. Finally, in this appendix subscripts denotes partial derivatives.

\subsection{Fully informed investors}\label{appendix-informed}

Assume that at time $0$  the agent observes $\Theta=\theta$, $\theta \in \mathbb R$, then he/she follows the standard Merton's investment rule
\begin{equation}
\pi ^{\ast}(t,\theta)=\frac{\theta}{\sigma \gamma},\;\;\; 0 \leq t \leq T,
\end{equation}
and enjoys an expected utility (cf. Rogers \cite{rogers-01}, Section 6)
\begin{equation}         %%%% Coincide con la (2.7) di Rogers
v^{\ast}(x;\theta) = \frac{x^{1 - \gamma}}{1 - \gamma}\exp\left(r(1 - \gamma)T+\frac{1  - \gamma}{2\gamma}\theta^2T\right).
\end{equation}
The expected utility (\ref{utilita-indiretta-informed-power}) for $\gamma \neq 1$ is obtained by integrating over $\theta$, that is
\begin{equation}
V^{I}(x;\gamma, T)= \int_{- \infty}^{+ \infty}\,v^{\ast}(x;\theta)p(\theta)d \theta,
\end{equation}
where $p$ denotes the Gaussian density with mean $\theta_0$ and variance $v_0$.

\subsection{Rational investors}\label{appendix-rational}
The optimal investment policy (\ref{optimal-portfolio-learner-power}) and expected utility (\ref{utilita-indiretta-learner-power}) are derived by using dynamic programming techniques (cf., Karatzas and Zhao \cite{karatzas-zhao-01} or Rogers \cite{rogers-01}, Section 6). For each $0\leq t \leq T$, $x > 0$ and $y \in \mathbb R$, let (see (\ref{wealth2}) and (\ref{process_Y-2}))
\begin{equation} \label{dynamic-rational-appendix}
\left\{
\begin{array}{l}
dX_s = rX_sds + \sigma \pi\left(s\right) X_s(\hat{\Theta}\left(s,Y_s\right)ds + d\hat W_s), \;\; X_t =x, \;\; t\leq s \leq T,\smallskip\\
dY_s = \hat{\Theta}\left(s,Y_s\right)ds + d\hat W_s, \;\; Y_t =y, \;\; t\leq s \leq T,
\end{array}%
\right.
\end{equation}
and define the value function
\begin{equation}
v^{R}(t,x,y) : = \sup_{\pi \in \mathcal{A}_t}\frac{1}{1 - \gamma}\mathbb E\left[X_T^{1 - \gamma}\right],
\end{equation}
where $\mathcal{A}_t$ is the set of all $\left(\mathcal{F}^S_t\right)$-progressively measurable and integrable controls $\pi=\left(\pi(s)\right)_{s \in [t,T]}$ such that (\ref{dynamic-rational-appendix}) admits a unique solution.
$v^{R}$, under appropriate regularity conditions, satisfies the Hamilton-Jacobi-Bellmann (HJB) equation
\begin{equation} \label{HJB-rational}
\sup_{\pi \in \mathbb R} \{v_t + rxv_x + \pi\sigma x\hat{\Theta}\left(t,y\right)v_x +\hat{\Theta}\left(t,y\right)v_y + \frac{1}{2}\pi^2\sigma^2x^2v_{xx} + \pi\sigma xv_{xy} + \frac{1}{2}v_{yy}\} = 0,
\end{equation}
for all $0 \leq t < T$, $x > 0$, $y \in \mathbb R$ and with boundary condition $v(T,x,y) = x^{1-\gamma}/(1-\gamma)$. Assuming $v_{xx} <0$, the maximization on the LHS of (\ref{HJB-rational}) yields the following candidate for the optimal (Markov) investment strategy:
\begin{equation} \label{portfolio-rational-appendix}
\pi^{\ast}(t,x,y) = - \frac{\hat{\Theta}(t,y)v_x }{\sigma x v_{xx}} - \frac{ v_{xy}}{\sigma x v_{xx}}.
\end{equation}
By substituting (\ref{portfolio-rational-appendix}) into HJB (\ref{HJB-rational}), the latter reduces to:
\begin{equation} \label{HJB-rational-reduced}
v_t + rxv_x - \frac{(\hat{\Theta}(t,y)v_x + v_{xy})^2}{2v_{xx}} + \hat{\Theta}(t,y) v_y  + \frac{1}{2}v_{yy} = 0,
\end{equation}
with the same boundary condition.
For $\gamma \neq 1$, that is, the non-logarithmic case,\footnote{For the logarithmic investor one proceeds similarly, albeit with a different trial function.} we try solutions of the form:
\begin{equation} \label{v-trial-rational-appendix}
v(t,x,y) = \frac{(xe^{r(T - t)})^{1 - \gamma}}{1 - \gamma}h(t,y),
\end{equation}
\begin{equation}
h(t,y) = \exp (a(t)\hat{\Theta}^2(t,y) + b(t)\hat{\Theta}(t,y) + c(t) ),
\end{equation}
\begin{equation}
a(T) = b(T) = c(T) = 0.
\end{equation}
The partial derivatives of $v$ are:
\begin{align*}
\smallskip
& v_t = \left( \frac{h_t}{h} - r(1 - \gamma) \right)v,\hspace{1.5cm} v_x = \frac{1  - \gamma}{x}v,\hspace{2cm} v_y = \frac{h_y}{h}v,\\
\smallskip
& v_{xx} = - \frac{\gamma(1  - \gamma)}{x^2}v,\hspace{2.7cm} v_{xy} = \frac{1  - \gamma}{x}\frac{h_y}{h}v, \hspace{1.5cm} v_{yy} = \frac{h_{yy}}{h}v.
\end{align*}
Notice that any solution of the form (\ref{v-trial-rational-appendix}) is such that $v_{xx}<0$. Substitute the previous derivatives into (\ref{HJB-rational-reduced}) and get the following PDE for $h$:
\begin{equation} \label{h-pde-rational}
\frac{h_t}{h}  + \frac{1}{\gamma}\hat{\Theta}(t,y)\frac{h_y}{h} + \frac{1 - \gamma}{2\gamma}\frac{h_y^2}{h^2}  + \frac{h_{yy}}{2h}
+ \frac{1 - \gamma}{2\gamma}\hat{\Theta}^2(t,y)= 0.
\end{equation}
The partial derivatives of $h$ can be written as follows:
\begin{align*}
\smallskip
& h_t = (a'\hat{\Theta}^2  + b'\hat{\Theta} + c' - (2a\hat{\Theta} + b)\hat{\Theta}\hat{\Theta}_y)h,\\
\smallskip
& h_y = (2a\hat{\Theta} + b)\hat{\Theta}_yh\\
\smallskip
& h_{yy} =((2a\hat{\Theta} + b)^2 + 2a)\hat{\Theta}_y^2h.
\end{align*}
where we use the identity $\hat{\Theta}_t = - \hat{\Theta}\hat{\Theta}_y$ (see definition (\ref{conditional-mean})).
By substituting into (\ref{h-pde-rational}) and collecting the powers of $\hat{\Theta}$, we have
\begin{eqnarray*}
0 &=&\left( a^{\prime }+\frac{2\hat{\Theta}_{y}^{2}}{\gamma }a^{2}+\frac{2\left( 1-\gamma \right) \hat{\Theta}_{y}}{\gamma }a+\frac{1-\gamma }{2\gamma }\right) \hat{\Theta}^{2} \\
  &+&\left( b^{\prime }+\frac{\left( 1-\gamma \right) \hat{\Theta}_{y}}{\gamma }b+\frac{2\hat{\Theta}_{y}^{2}}{\gamma }ab\right) \hat{\Theta} \\
  &+&c^{\prime }+\hat{\Theta}_{y}^{2} a+\frac{\hat{\Theta}_{y}^{2}}{2\gamma }b^{2}.
\end{eqnarray*}
Since $\hat{\Theta}_y =  v_0/(1 + v_0t)$ does not depend on $y$, the previous equality holds for all $0\leq t < T$ and $y \in \mathbb R$ if, and only if, functions $a(t)$, $b(t)$ and $c(t)$ solve, for all $0\leq t < T$, the system of ODEs:
\begin{equation}
\left\{
\begin{array}{l}
a^{\prime }+\dfrac{2\hat{\Theta}_{y}^{2}}{\gamma }a^{2}+\dfrac{2\left( 1-\gamma \right) \hat{\Theta}_{y}}{\gamma }a+\dfrac{1-\gamma }{2\gamma } \smallskip =0 \\
b^{\prime }+\dfrac{\left( 1-\gamma \right) \hat{\Theta}_{y}}{\gamma }b+\dfrac{2\hat{\Theta}_{y}^{2}}{\gamma }ab=  \smallskip 0 \\
c^{\prime }+\hat{\Theta}_{y}^{2}a+\dfrac{\hat{\Theta}_{y}^{2}}{2\gamma }b^{2}=0.%
\end{array}%
\right.
\end{equation}
The system is explicitly solvable (although with some algebra) and the particular solution that satisfies the terminal condition $a(T) = b(T) = c(T) = 0$ is:
\begin{align*}
\smallskip
& a^{R}(t) = \frac{(1 - \gamma)(1 + v_0t)(T - t)}{2(\gamma(1 + v_0T) - v_0T  + v_0t)},\\
\smallskip
& b^{R}(t) = 0,\\
\smallskip
& c^{R}(t) =  \frac{1}{2}\left(\gamma\ln\frac{\gamma(1 + v_0T)}{\gamma(1 + v_0T) - v_0T + v_0t} - \ln\frac{1 + v_0T}{1 + v_0t}\right),
\end{align*}
for each $0\leq t\leq T$. Now, a standard verification argument proves
\begin{equation}
v^{R}(t,x,y)=\frac{(xe^{r(T-t)})^{1-\gamma }}{1-\gamma }\exp (a^{R}(t)\hat{\Theta}^{2}(t,y)+b^{R}(t)\hat{\Theta}(t,y)+c^{R}(t)), \;\; \gamma \neq 1,
\end{equation}%
for all $0\leq t\leq T$, $x>0$ and $y\in \mathbb{R}$. The expected utility (\ref{utilita-indiretta-learner-power}) for $\gamma \neq 1$ is obtained by fixing $t=0$ and $y=0$ in $v^{R}$, that is%
\begin{equation}
V^{R}\left( x;\gamma ,T\right) =v^{R}(0,x,0)
\end{equation}
(notice that $\varphi ^{R}\left( \gamma ,T\right) =a^{R}\left( 0\right) $ and $\psi ^{R}\left( \gamma ,T\right) =c^{R}\left( 0\right)  +r\left( 1-\gamma \right) T$).
Moreover,
\begin{eqnarray*}
\pi ^{\ast }(t,x,y) &=&\frac{\hat{\Theta}(t,y)}{\sigma \gamma }+\frac{1}{%
\sigma \gamma }\frac{h_{y}}{h}=\frac{\hat{\Theta}(t,y)}{\sigma \gamma }+%
\frac{\hat{\Theta}(t,y)}{\sigma \gamma }\left( 2a^{R}\left( t\right) \hat{%
\Theta}_{y}(t,y)\right)  \\
&=&\frac{\hat{\Theta}(t,y)}{\sigma \gamma }+\frac{\hat{\Theta}(t,y)}{\sigma
\gamma }\left( \frac{(1-\gamma )(T-t)v_{0}}{\gamma (1+v_{0}T)-v_{0}T+v_{0}t}%
\right) ,
\end{eqnarray*}%
which, once substituted the optimal state dynamics, yields the optimal investment strategy (\ref{optimal-portfolio-learner-power}). That is,
$\pi _{\gamma }^{R}(t)=\pi ^{\ast }(t,X_{t},Y_{t})$.

\subsection{Myopic investor}\label{appendix-myopic}

Substitute the myopic investment strategy (\ref{optimal-portfolio-myopic-power}) into (\ref{wealth2}) to get the state dynamics
\begin{equation} \label{dynamic-myopic-appendix}
\left\{
\begin{array}{l}
dX_s = rX_sds +  \dfrac{\hat{\Theta}(s,Y_s)}{ \gamma } X_s(\hat{\Theta}\left(s,Y_s\right)ds + d\hat W_s), \;\; X_t =x, \;\; t\leq s \leq T,\smallskip\\
dY_s = \hat{\Theta}\left(s,Y_s\right)ds + d\hat W_s, \;\; Y_t =y, \;\; t\leq s \leq T.
\end{array}%
\right.
\end{equation}
Define
\begin{equation}
v^{M}(t,x,y) : = \frac{1}{1 - \gamma}\mathbb E\left[X_T^{1 - \gamma}\right].
\end{equation}
Then, under appropriate regularity conditions, $v^{M}$ satisfies the PDE%\footnote{Subscripts denote partial derivatives.}
\begin{equation} \label{PDE-myopic}
v_t + rxv_x + \frac{1}{\gamma}\hat{\Theta}^{2}(t,y)xv_x + \hat{\Theta}(t,y)v_y + \frac{1}{2\gamma^2}\hat{\Theta}^{2}(t,y)x^2v_{xx}
+ \frac{1}{\gamma}\hat{\Theta}(t,y)xv_{xy} + \frac{1}{2}v_{yy} = 0,
\end{equation}
for all $0 \leq t < T$, $x > 0$, $y \in \mathbb R$ and with boundary condition $v(T,x,y) = x^{1-\gamma}/(1-\gamma)$.
For $\gamma \neq 1$, we proceed similarly to the rational case and try solutions of the form:
\begin{equation} \label{v-trial-myopic-appendix}
v(t,x,y) = \frac{(xe^{r(T - t)})^{1 - \gamma}}{1 - \gamma}h(t,y),
\end{equation}
\begin{equation}%
h(t,y) = \exp (a(t)\hat{\Theta}^2(t,y) + b(t)\hat{\Theta}(t,y) + c(t) ),
\end{equation}
\begin{equation}
a(T) = b(T) = c(T) = 0.
\end{equation}
The partial derivatives of $v$ are as in the rational case and, once substituted into (\ref{PDE-myopic}),  yield the following PDE for $h$:
\begin{equation} \label{h-pde-myopic}
\frac{h_t}{h}  + \frac{1}{\gamma}\hat{\Theta}(t,y)\frac{h_y}{h} + \frac{h_{yy}}{2h} + \frac{1 - \gamma}{2\gamma}\hat{\Theta}^2(t,y) = 0.
\end{equation}
Again, the partial derivatives of $h$ are as before and (\ref{h-pde-myopic}) becomes:
\begin{eqnarray*}
0 &=&\left( a^{\prime }+ 2\hat{\Theta}_{y}^{2}a^{2}+\dfrac{2\left( 1-\gamma \right) \hat{\Theta}_{y}}{\gamma }a+\dfrac{1-\gamma }{2\gamma } \right) \hat{\Theta}^{2} \\
  &+&\left( b^{\prime }+\dfrac{\left( 1-\gamma \right) \hat{\Theta}_{y}}{\gamma }b+2\hat{\Theta}_{y}^{2}ab \right) \hat{\Theta} \\
  &+&c^{\prime }+\hat{\Theta}_{y}^{2}a+\dfrac{\hat{\Theta}_{y}^{2}}{2}b^{2},
\end{eqnarray*}
which holds for all $0\leq t < T$ and $y \in \mathbb R$ if, and only if, functions $a(t)$, $b(t)$ and $c(t)$ solve, for all $0\leq t < T$, the system of ODEs:
\begin{equation}
\left\{
\begin{array}{l}
a^{\prime }+ 2\hat{\Theta}_{y}^{2}a^{2}+\dfrac{2\left( 1-\gamma \right) \hat{\Theta}_{y}}{\gamma }a+\dfrac{1-\gamma }{2\gamma } \smallskip =0 \\
b^{\prime }+\dfrac{\left( 1-\gamma \right) \hat{\Theta}_{y}}{\gamma }b+2\hat{\Theta}_{y}^{2}ab =  \smallskip 0 \\
c^{\prime }+\hat{\Theta}_{y}^{2}a+\dfrac{\hat{\Theta}_{y}^{2}}{2}b^{2}=0.
\end{array}%
\right.
\end{equation}
The system is explicitly solvable and, after some algebra, the solution that satisfies the terminal condition $a(T) = b(T) = c(T) = 0$ is:\footnote{Notice that the equation for $a$ together with $a\left( T\right) =0$ enable to prove that $a^{M}\left( t\right) <0$ ($a^{M}\left( t\right) >0$), $0\leq t<T$, for $\gamma >1$ ($0<\gamma <1$), hence $%
r_{2}(1+v_{0}T)^{r_{2}-r_{1}}-r_{1}(1+v_{0}t)^{r_{2}-r_{1}}>0$ for all $ 0\leq t\leq T$. Moreover, the equation for $c$ together with $c^{M}\left( T\right) =0$ and the fact that $b^{M}\left( t\right) =0$ imply that $ c^{M}\left( t\right) <0$ ($c^{M}\left( t\right) >0$), $0\leq t<T$, and increasing (decreasing) for $\gamma >1$ ($0<\gamma <1$).}
\begin{align*}
\smallskip
& a^{M}(t)=\frac{(1-\gamma )\left( 1+v_{0}t\right) \left( (1+v_{0}T)^{r_2 - r_1}-(1+v_{0}t)^{r_2 - r_1}\right) }{2\gamma
v_{0}\left( r_{2}(1+v_{0}T)^{r_2 - r_1}-r_{1}(1+v_{0}t)^{r_2 - r_1}\right) },\\
\smallskip
& b^{M}(t) = 0,\\
\smallskip
& c^{M}(t)=\frac{1}{2}\left( \ln \frac{r_{2}-r_{1}}{r_{2}(1+v_{0}T)^{r_{2}-r_{1}}-r_{1}(1+v_{0}t)^{r_{2}-r_{1}}}+\ln \frac{(1+v_{0}T)^{r_{2}}}{(1+v_{0}t)^{r_{1}}}\right),
\end{align*}
where $r_1 < r_2$ are the roots of $r^2 + (2/\gamma -1  )r + (1 - \gamma)/\gamma$.
Hence,
\begin{equation}
v^{M}(t,x,y)=\frac{(xe^{r(T-t)})^{1-\gamma }}{1-\gamma }\exp (a^{M}(t)\hat{\Theta}^{2}(t,y)+b^{M}(t)\hat{\Theta}(t,y)+c^{M}(t)), \;\; \gamma \neq 1,
\end{equation}%
for all $0\leq t\leq T$, $x>0$, $y\in \mathbb{R}$, and the expected utility (\ref{utilita-indiretta-myopic-power}), for $\gamma \neq 1$, is obtained by fixing $t=0$ and $y=0$ in $v^{M}$, that is
\begin{equation}
V^{M}\left( x;\gamma ,T\right) =v^{M}(0,x,0)
\end{equation}
(notice that $\varphi ^{M}\left( \gamma ,T\right) =a^{M}\left( 0\right) $ and $\psi ^{M}\left( \gamma ,T\right) =c^{M}\left( 0\right)  +r\left( 1-\gamma \right) T$).

\subsection{Unconditional investors}\label{appendix-unconditional}

Substitute the myopic investment strategy (\ref{optimal-portfolio-static-power}) into (\ref{wealth2}) to get the state dynamics
\begin{equation} \label{dynamic-static-appendix}
\left\{
\begin{array}{l}
dX_s = rX_sds +  \dfrac{\theta_0}{ \gamma } X_s(\hat{\Theta}\left(s,Y_s\right)ds + d\hat W_s), \;\; X_t =x, \;\; t\leq s \leq T,\smallskip\\
dY_s = \hat{\Theta}\left(s,Y_s\right)ds + d\hat W_s, \;\; Y_t =y, \;\; t\leq s \leq T.
\end{array}%
\right.
\end{equation}
Define
\begin{equation}
v^{U}(t,x,y) : = \frac{1}{1 - \gamma}\mathbb E\left[X_T^{1 - \gamma}\right].
\end{equation}
Then, under appropriate regularity conditions, $v^{U}$ satisfies the PDE:%\footnote{Subscripts denote partial derivatives.}
\begin{equation} \label{PDE-static}
v_t + rxv_x + \frac{1}{\gamma}\hat{\Theta}(t,y)\theta_0 xv_x + \hat{\Theta}(t,y)v_y + \frac{\theta_0^2}{2\gamma^2}x^2v_{xx}
+ \frac{\theta_0}{\gamma}xv_{xy} + \frac{1}{2}v_{yy} = 0,
\end{equation}
$0 \leq t < T$, $x > 0$, $y \in \mathbb R$, with boundary condition $v(T,x,y) = x^{1-\gamma}/(1-\gamma)$.
For $\gamma \neq 1$, we proceed as in the previous two cases and try solutions of the form:
\begin{equation} \label{v-trial-static-appendix}
v(t,x,y) = \frac{(xe^{r(T - t)})^{1 - \gamma}}{1 - \gamma} \exp (a(t)\hat{\Theta}(t,y) + b(t) ),
\end{equation}
\begin{equation}
a(T) = b(T)= 0.
\end{equation}
Computations similar to the previous two cases proves that $v$ solves (\ref{PDE-static}) if, and only if, the functions $a(t)$ and $b(t)$ solve, for all $0\leq t < T$, the system of ODEs:
\begin{equation}
\left\{
\begin{array}{l}
a^{\prime }+ \dfrac{1-\gamma }{\gamma }\theta_0 \smallskip =0 \\
b^{\prime }+\dfrac{1-\gamma }{\gamma }\theta _{0}\hat{\Theta}_{y}a+\dfrac{1}{2}\hat{\Theta}_{y}^{2}a^{2}-\dfrac{1-\gamma }{2\gamma }\theta _{0}^{2} = 0 .
\end{array}%
\right.
\end{equation}
The solution that satisfies the terminal condition $a(T) = b(T)= 0$ is:
\begin{align*}
\smallskip
& a^{U}(t)=\frac{1 - \gamma}{\gamma}\theta_0(T - t),\\
\smallskip
& b^{U}(t) = \frac{(1 - \gamma)(T - t)}{2\gamma^2}\left((1 - \gamma)\frac{1 + v_0T}{1 + v_0t} - 1\right)\theta_0^2,
\end{align*}
$0\leq t\leq T$.
Hence,
\begin{equation}
v^{U}(t,x,y)=\frac{(xe^{r(T-t)})^{1-\gamma }}{1-\gamma }\exp (a^{U}(t)\hat{\Theta}(t,y)+b^{U}(t)), \;\; \gamma \neq 1,
\end{equation}%
for all $0\leq t\leq T$, $x>0$, $y\in \mathbb{R}$, and the expected utility (\ref{utilita-indiretta-static-power}), for $\gamma \neq 1$, is obtained by fixing $t=0$ and $y=0$ in $v^{U}$, that is
\begin{equation}
V^{U}\left( x;\gamma ,T\right) =v^{U}(0,x,0).
\end{equation}

\end{document}